\documentclass[showpacs,twocolumn,aps,preprintnumbers,letterpaper]{revtex4}
\usepackage{amsmath,amssymb}
\usepackage{epsfig}
\usepackage{graphicx}
\usepackage{amsmath}
\usepackage{slashed}
\usepackage{amsfonts}
\usepackage{epstopdf}
\usepackage{color}

\newcommand{\be}{\begin{equation}}
\newcommand{\ee}{\end{equation}}
\newcommand{\bear}{\begin{eqnarray}}
\newcommand{\eear}{\end{eqnarray}}
\newcommand{\ba}{\begin{array}}
\newcommand{\ea}{\end{array}}

\def\be{\begin{eqnarray}}
\def\ee{\end{eqnarray}}
\def\bea{\be}
\def\eea{\ee}

\newcommand{\e}{{\mbox{e}}}

\def\roughly#1{\mathrel{\raise.3ex\hbox{$#1$\kern-.75em%
\lower1ex\hbox{$\sim$}}}}

\begin{document}

\title{Hydrodynamical corrections to electromagnetic emissivities in QCD}

\author{Yizhuang Liu  and Ismail Zahed}
\email{yizhuang.liu@stonybrook.edu}
\email{ ismail.zahed@stonybrook.edu}
\affiliation{Department of Physics and Astronomy, Stony Brook University, Stony Brook, New York 11794-3800, USA}


\date{\today}
\begin{abstract}
We provide a general framework for the derivation of the hydrodynamical corrections to the QCD electromagnetic 
emissivities in a viscous fluid. Assuming that the emission times are short in comparison to the fluid evolution
time, we show that the leading corrections in the fluid gradients are controlled by the bulk and shear tensors times 
pertinent response functions  involving the energy-momentum tensor. In a hadronic fluid phase, we explicit these
contributions using spectral functions. Using the vector dominance approximation, we show that 
the bulk viscosity correction to the photon rate is sizable, while the shear viscosity is negligible for about all frequencies.  
In the partonic phase near the transition temperature we provide an assessment
of the viscous corrections to the photon and dilepton emissions,
using a non-perturbative quark-gluon plasma with soft thermal gluonic corrections in the form of 
operators of leading mass dimension. Again, the thermal bulk viscosity corrections  are found to be larger than
the thermal shear viscosity corrections at all energies for both the photon and dilepton in the partonic phase.

 \end{abstract}


\maketitle

\setcounter{footnote}{0}


\section{Introduction}

One of the major achievement of the heavy ion program at RHIC and now also at LHC is
the emergence of a new state of  matter under extreme conditions, 
the strongly coupled quark gluon plasma (sQGP) with 
near ideal liquid properties~\cite{Heinz:2013th,Luzum:2013yya,Teaney:2009qa}.  The prompt
release of a large entropy in the early partonic phase together with a rapid thermalization and 
short mean free paths, points to  a partonic  fluid. The anisotropies of the produced
hadrons and  photons suggest a near ideal fluid~\cite{Adl03,PHE12,PHE16,ALICE12,ALICE13,CMS13,CMS14}.

Small deviations from the ideal limit appears to follow from dissipative effects, suggesting that
the shear viscosity of the sQGP fluid is very  close to its quantum bound~\cite{SON}. However, this
interpretation requires some care since the emitted hadrons interact strongly throughout the 
fluid hystory,  and particularly in the late stages of the evolution composed essentially of a
 fluid of hadrons. In  contrast, the emitted  photons or dileptons are continuously emitted 
 throughout the evolution of the fluid without secondary interactions. They provide for an 
alternative probe of the nature and strength of these  viscous corrections.

 In so far, most of the hydrodynamical corrections to the electromagnetic emissivities have
 made use of weakly coupled kinetic theory to modify the phase space distributions of either
 partons or hadrons in $2\rightarrow 2$ rate processes~\cite{VISCO}. 
 Holographic calculations for the electromagnetic emissivities  for  
 ${\cal N}=4$ SUSY were carried in near equilibrium in~\cite{HONG}, and far from 
 equilibrium in~\cite{BAIER}. In light of this, 
  It is important to seek a full non-perturbative analysis of the
 electromagnetic emissivities in a viscous QCD fluid that  relies solely on a near-equilibrium 
 approximation and a  fluid gradient expansion.

 The purpose of this paper is to provide such a framework for the analysis of the emission
 of photons and dileptons from a non-ideal hydrodynamical QCD fluid that does not rely on 
 perturbation theory. Assuming that the electromagnetic emission time is shorter than the 
 fluid unfolding time, we show how to organize the rates in the near equilibrium phase by 
 expanding in the fluid derivatives. The emerging fluid bulk and shear tensors are multiplied by pertinent 
 correlation functions involving the energy-momentum tensor in equilibrium. 
 
 The organization of the paper is as follows: in section 2 we show how to assess the electromagnetic
 emissivities in a fluid near equilibrium by capturing the slow fluid flow in a density matrix. In section 3,
 we show that in leading order in the fluid gradients, the electromagnetic emissivities receive
 contributions proportional to the bulk and shear tensors times Kubo-like response functions involving the
 energy momentum tensor. 
 In section 4, we analyze the leading viscous corrections to the electromagnetic emissivities in the hadronic
 phase, and in section 5 in a non-perturbative partonic phase. Our conclusions are in section 6. A background field 
 analysis for the soft gluon corrections in the partonic phase is outlined in Appendix A. We also detail the leading
 contribution to the photon thermal viscous corrections in Appendix B.

\section{Photon emission in a fluid}

In thermal equilibrium, the photon  emission rate is fixed by the Wightman function for the electromagnetic current
\cite{LARRY}

\be
\label{1}
\frac{d\Gamma_0}{d^3k}=-\frac{\alpha\, g^{\mu\nu}}{4\pi^2|k|}\left<\mathbb G_{\mu\nu}^{<}(q)\right>_0
\ee
with

\be
\label{2}
\mathbb G_{\mu\nu}^{<}(q)=\int d^4x e^{-iq\cdot x}\,J_{\mu}(0)J_{\nu}(x)
\ee
The averaging is carried over the state of maximum entropy or equivalently a thermal distribution of fixed
temperature $1/\beta$. In writing (\ref{2}) space-time translational invariance is assumed. Most studies of 
photon emission at collider energies have relied on (\ref{1}), with some recent exceptions using modifications based
on kinetic theory.

For a system far out of equilibrium  its evolution and emission rates are convoluted. However,  for large times
the system nears equilibrium and its evolution follows the lore of hydrodynamics. In this regime, the microscopic
electromagnetic emission rates can be assumed to occur on time scales shorter than the times it takes for the
fluid to flow. In this decoupling approximation, we may ask for the changes caused by a fluid velocity profile on
the electromagnetic emissivities of a QCD fluid for instance.

With this in mind, we may still rely on (\ref{2}) at 
any time $\bar t$ since space-time microscopic translational invariance holds. Now, 
consider the emission on a fluid time-like surface defined by $\bar t={\rm constant}$ and canonically quantize the field
theory on this surface. Let $\phi_{\bar t}$ be a generic operator on this surface. Its time evolution proceeds through

\be
\phi_{\bar t}(t,\vec{x})\equiv \phi(\bar t +t,\vec x)=e^{iHt}\phi(\bar t,\vec x)e^{-iHt}
\ee
with the canonical Hamiltonian $H=H[\phi_{\bar t},\pi_{\bar t}]$. The  emission on this time-like surface is still controlled by the
general Wightman  function

\be
\label{3}
\left<\mathbb G^<_{\bar t\,\mu\nu}(t,\vec{x})\right>= {\rm Tr}\left(\rho(\bar t_0)\,J_{\bar t\mu}(0,\vec{0})J_{\bar t\nu}(t,{\vec{x}})\right)
\ee
with an initial density operator at $\bar t_0<\bar t$.  For a state in equilibrium, we have

\be
\rho(\bar t_0)\rightarrow \rho(\bar t)=e^{-\beta_{\bar t}(H-F_{\bar t})}
\ee
However, for a state near-equilibrium we define

\be
\label{4}
\rho(\bar t_0)=\rho(\bar t)U(\bar t,\bar t_0)\equiv 
\rho(\bar t)\,T_\tau e^{\int_{0}^{1}d\tau \Sigma(\bar t-i\beta_{\bar t}\tau,\bar t_0)}
\ee
The operator $\Sigma$ is a measure of the entropy change from $\bar t_0\rightarrow \bar t$ as discussed in~\cite{OUT}.
For our case, it is sufficient to note that it follows from the covariatized gradient expansion of $\beta H$,

\be
\label{4X}
\Sigma(\bar t-i\beta\tau,\bar t_0)=\partial_{i}\beta_j\int_{\bar t_0}^{\bar t}dt^{\prime}\int_{0}^1d\tau  d^3x^\prime\,T_{ij}(\vec{x}{\,^\prime} ,t^{\prime}-i\tau\beta_{\bar t})\nonumber\\
\ee
for a time-independent but spatially dependent fluid velocity $\beta_i$. 
We note that (\ref{4}) can be equally defined through

\be
\label{4X}
\rho(\bar t_0)=\tilde U(\bar t,\bar t_0)\rho(\bar t)\equiv T^{\star}_\tau e^{\int_{0}^{1}d\tau \Sigma(\bar t+i\beta_{\bar t} \tau,\bar t_0)}\rho(\bar t)
\ee
 For $\bar t\gg \bar t_0$, the averaging over $\rho(\bar t_0)$ asymptotes the
equilibrium average captured by $\rho(\bar t)$, modulo derivative corrections 
due to the fluid gradients as captured in $\Sigma$. 
In what will follow, we will set $\beta=\beta_{\bar t}$ for notational simplicity.

\section{Gradient expansion}

For a baryon free fluid flow characterized locally by $\beta_i$, we  can now organize (\ref{3}) using an expansion in fluid gradients $\partial_i\beta_j$. For a given time $\bar t$, the leading contribution emerges only by keeping 
$\rho(\bar t)$ in (\ref{4}). In this order (\ref{3}) yields  (\ref{1}) in equilibrium. The fluid gradient corrections appear at next to leading order by expanding the $\tau$-ordered exponent and retaining only the first gradient correction in $\Sigma$,

\bea
\label{4XX}
&&\rho (\bar t_0)\\
&&\approx\rho(\bar t)
\left(1-\partial_{i}\beta_j\int_{\bar t_0}^{\bar t}dt^{\prime}\int_{0}^1d\tau  d^3x^\prime\,T_{ij}(\vec{x}{\,^\prime} ,t^{\prime}-i\tau\beta)\right)\nonumber \\ 
&&\approx \left(1-\partial_{i}\beta_j\int_{\bar t_0}^{\bar t}dt^{\prime}\int_{0}^1d\tau  d^3x^\prime\,  T_{ij}(\vec{x}{\,^\prime} ,t^{\prime}+i\tau\beta)\right)\rho (\bar t)\nonumber
\eea
Inserting (\ref{4XX}) in (\ref{3}) yields the first order fluid gradient correction to the electromagnetic emissivities

\bea
\label{5}
&&\left<\mathbb G_{\bar t \, \mu\nu}^{<}(t,\vec x)\right>_1\approx\nonumber \\ 
&&-\int_{\bar t_0}^{\bar t} dt^{\prime}\int_0^1d\tau\left<T_{ii}(t^{\prime}-i\tau\beta, \vec q=0)J_{{\bar t\mu}}(0)J_{\bar t\nu}(x)\right>_{\beta}\,\theta\nonumber \\ 
&&-\int_{\bar t_0}^{\bar t} dt^{\prime}\int_0^1d\tau\left<T_{ij}(t^{\prime}-i\tau\beta,\vec q=0)J_{{\bar t \mu}}(0)J_{\bar t \nu}(x)\right>_{\beta}\sigma_{ij}\nonumber\\
\eea
with $\theta={\partial_m\beta_m}/3$. 
The transverse and traceless shear velocity tensor is defined as 

\be
\sigma_{ij}=\frac{1}{2}\left(\partial_i\beta_j+\partial_j\beta_i-\frac{2}{3}\delta_{ij}\,\partial_m\beta_m\right)
\ee
 (\ref{5}) involves the causal change in the electromagnetic emissivity caused by the fluid bulk and shear parts of the energy
momentum tensor $T_{ij}$, while evolving from $\bar t_0\rightarrow \bar t$. 

The Kubo-like 3-point response function in (\ref{5}) can be made more explicit by defining

\be
\label{5X1}
 O_{ij}^{\pm}=\int_{\bar t_0}^{\bar t} dt^{\prime}\int_0^1d\tau \int d^3 x^\prime T_{ij}(t^{\prime}\mp i\tau\beta, \vec x^\prime )
\ee
so that 

\bea
\label{5X11X}
&&\left<\mathbb G_{\bar t \, \mu\nu}^{<}(t,\vec x)\right>_1\approx\nonumber \\ 
&&-\left<O_{ii}^-J_{{\bar t \mu}}(0)J_{\bar t \nu}(x)\right>\,\theta
-\left<O_{ij}^-J_{{\bar t \mu}}(0)J_{\bar t \nu}(x)\right>\sigma_{ij}\nonumber\\
\eea
The equivalence between the left-right decomposition in (\ref{4XX}) suggests that the operator
$O_{ij}^\pm$ commutes with the Hamiltonian. Indeed, we have 

\bea
\label{5X2}
[H, O_{ij}^{\pm}]=-i\int_{\bar t_0}^{\bar t} dt^{\prime}\int_0^1d\tau \int d^3 x^\prime \partial_{t^{\prime}}T_{ij}(t^{\prime}\mp i\tau\beta, \vec x^\prime)\nonumber \\ =\pm\frac{1}{\beta}\int_{\bar t_0}^{\bar t} dt^{\prime}d^3 x^\prime (T_{ij}(t\mp i\beta,\vec{x})-T_{ij}(t,x^\prime))
\eea
If the de-correlation time in the Kubo-like result (\ref{5X11X}) is short in comparison to the fluid evolution time, 
we may regard $t_H=\bar t-\bar{t}_0$ as large. This will be understood throughout.
Therefore, the commutator in (\ref{5X2}) vanishes modulo asymptotic terms. 
We note that the operator $O_{ij}$ is related to the time integration of the first moment
of the momentum density which is conserved,

\bea
&&\partial_t\int d^3x\, (x_iT^0_{j}(t,\vec{x}))=\int d^3x\, (x_i\partial_tT^0_{j}(t,\vec{x}))\nonumber \\ 
&&=-\int d^3x (x_i\partial_kT^k_{j}(t,\vec{x}))=\int d^3x\, T_{ij}(t,\vec{x})
\eea
It follows that its expectation value is proportional to the time length $t_H$ characteristic of the hydrodynamical
evolution, which is assumed to be much larger than the characteristic time for electromagnetic emission. This point
will become clear in the explicit calculations to follow.

\section{Hadronic phase}

In a QCD fluid the analysis of the response functions depend on the nature of the underlying phase.
At low temperatures, the fluid is mostly hadronic, while at high temperature it is partonic but strongly coupled
near the cross-over temperature. In the hadronic phase  with zero baryon density and no strangeness, 
(\ref{5X11X}) can be organized by expanding it in  increasing densities of the lightest stable thermal hadrons, i.e. pions 
following  similar analyses for the equilibrium rates in~\cite{STEELE}. Specifically, we have

\bea
\label{5X8}
&&G^<_{\bar t ij}(x)={\rm Tr}\left(e^{-\beta (H-F)}O^\pm_{ij}\,\left(J(\bar x)J(\bar x-x)\right)\right)=\nonumber\\
&&G^<_{ij,0\pi}+\int (d\tilde \pi_1)_{ij}\,G^<_{1\pi}+\frac 1{2!}\int (d\tilde \pi_1d\tilde \pi_2)_{ij}\, G^<_{2\pi}+...
\eea
where we have defined the unordered and connected  matrix elements

\bea
\label{5X9}
&&G^<_{n\pi}(x)=\nonumber\\
&&\left<\pi^{a_1}(k_1)...\pi^{a_n}(k_n)|J(\bar x)J(\bar x-x)|\pi^{a_1}(k_1)...\pi^{a_n}(k_n)\right>\nonumber\\
\eea
with the pion thermal phase space factors ($E_i^2=\vec k_i^2+m_\pi^2$)

\be
\label{5X10}
(d\tilde \pi_1)_{ij}=\frac{d^3k_i}{(2\pi)^3}\frac {t_H}{2E_i}\frac{k_ik_j}{E_i}n_{B}(E_i)(1+n_B(E_i))
\ee
and the identification $t_H=2\pi\delta(0_E)$.
This can be justified by explicitly performing the trace using the in-states. For instance for the 
1-pion connected pieces, we have 

\bea
&&\sum_{n, [k]} \left<\pi (k_1)..\pi(k_n)|e^{-\beta H}O^\pm_{ij}JJ|\pi(k_1)..\pi(k_n)\right>=\nonumber \\ 
&&\sum_{n_2, k_2, ...}...\sum_{n1, k_1} \frac{n_1^2}{n_1}
\left<\pi(k_1)|JJ|\pi(k_1)\right>e^{-n_1\beta E_{k_1}}\frac{k_{1i}k_{2j}}{E_1}=\nonumber \\
 && e^{-\beta F_0}\int \frac{d^3k}{(2\pi)^3}\frac {t_H}{2E_k}\frac{k_ik_j}{E_k}n_B(1+n_B)\left<\pi(k)|JJ|\pi(k)\right>\nonumber\\
\eea

\subsection{$G^<_{ij,0\pi}$ contribution}

The contributions to $G^<_{ij,0\pi}$ in (\ref{5X8}) follow from $2\pi, 4\pi, ...$ insertions in the intermediate state, 
and are found to all vanish. Indeed, consider the leading $2\pi$ insertion to $G^<_{ij,0\pi}$

\bea
\label{5X11}
&&G^<_{ij,0\pi}(x)=\left<0|O^\pm_{ij} J(\bar x)J(\bar x-x)|0\right>\approx \nonumber\\
&&2\int \frac{d^3k_1d^3k_2}{(2\pi)^62E_{k1}2E_{k_2}} \left<0|O_{ij}^{\pm}|\pi^+(k_1)\pi^-(k_2)\right>\nonumber\\
&&\times \left<\pi^+(k_1)\pi^-(k_2)|J(\bar x)J(\bar x-x)|0\right>\nonumber\\
\eea
where the overall factor 2 accounts for isospin. 
 The covariantize  transition matrix element in (\ref{5X11})  in leading order in the pion momentum reads

\bea
\label{5X12}
&&\left<0|T_{\mu \nu}(t,x)|\pi^+(k_1)\pi^-(k_2)\right>=\nonumber\\
&& e^{-it(E_1+E_2)+i\vec{x}\cdot (\vec{k1}+\vec{k_2})}\nonumber\\&& 
\times(-k_{1{\mu}}k_{2{\nu}}-k_{2{\mu}}k_{1{\nu}}-g_{\mu\nu}(-k_1\cdot k_2-m_\pi^2))
\eea
At asymptotic times or $t$ large as required by the out-field condition, this contribution vanishes
owing to the non-vanishing Fourier component in time. This result is consistent with the
fact that $O_{mn}$ connects only states with $E_m=E_n$. Clearly, this result carries to all $2n\pi$ insertions,
making $G^<_{ij,0\pi}=0$.

\subsection{$G^<_{1\pi}$ contribution}

The leading correction to (\ref{5X11X}) arises from the thermal one-pion contribution to $G^<_{1\pi}$.
Specifically, we have

\bea
\label{5X15Z1}
(d\tilde\pi_1)_{ij}\, G^<_{1\pi}(x)&&=\frac{d^3k_1}{(2\pi)^3}\frac{t_H}{2E_1}\frac{k_{1i}k_{1j}}{E_1}n_B(1+n_B)\nonumber\\
&&\times\left<\pi^a(k_1)|J(\bar x)J(\bar x-x)|\pi^a(k_1)\right>
\eea
which is seen to involve part of the forward photon-pion scattering amplitude. 
 Its explicit form follows from the general strictures of 
broken chiral symmetry, crossing symmetry and unitarity~\cite{YAMA,STEELE},

\bea
\label{5X16}
( d\tilde\pi_1)_{ij}\, G^<_{1\pi}(q)=&&\frac{d^3k_1}{(2\pi)^3}\frac{t_H}{2E_1}\frac{k_{1i}k_{1j}}{E_1}n_B(1+n_B)
\nonumber\\&&\times\biggl(
-\frac{6}{f_\pi^2}(k_1-q)^2\text{Im} {\bf \Pi}_A \left( (k_1-q)^2\right) \biggr)\nonumber\\
\eea
Here $\Pi_A$ is the $AA$ correlation function of the axial-vector current in the vacuum~\cite{STEELE}. 
Its spectral form follows from $\tau$-decay measurements into an odd number of pions. The result
(\ref{5X16}) grows linearly with the  hydrodynamical time $t_H$, that is the time it takes 
the externally applied hydrodynamical gradient $\partial_i\beta_j$  to change. This time 
is proportional to the transport mean free path $t_H\approx \lambda_{\rm mfp}$, 
which in turn is  determined by the viscosities,

\bea
\label{HYDROTIME}
&&t_H\rightarrow t_\eta\approx \frac{\eta}{e+p}\qquad {\rm shear}\nonumber\\
&&t_H\rightarrow t_\zeta\approx \frac{\zeta}{e+p}\qquad {\rm bulk}
\eea
Here $\eta, \zeta$ are the shear and bulk viscosities respectively, and $e, p$ are the energy and pressure densities
respectively. These hydrodynamical times will be understood in the results to follow.

\subsection{Viscous photon rate}

The viscous corrections to the photon rates due to a baryon free fluid of hadrons follow from the results in
(\ref{1}-\ref{5X11X}) and in (\ref{5X8}-\ref{5X16}), in the form

\bea
\label{Z1XX}
\frac{d\Gamma^<_1}{d^3k}=&&- \frac{\alpha\,t_H}{2\pi^2\omega}\int \frac {d^3p}{(2\pi)^3}\frac 1{2E} 
\frac {e^{\beta E}}{(e^{\beta E}-1)^2}\,\frac {p^2}{E}\nonumber\\
&&\times\left(2\partial_{i}\beta_{j}{\bf P}_{ij}\cos^2 \theta_p+\partial_{i}\beta_{j}{\bf N}_{ij} \sin^2 \theta_p\right)\nonumber\\
&&\times \left(-\frac{3}{f_\pi^2}(p-k)^2\text{Im} {\bf \Pi}_A \left( (p-k)^2\right) \right)\nonumber\\
\eea
with ${\bf P}_{ij}=\hat k_i \hat k_j$ and ${\bf N}_{ij}=\delta_{ij}-\hat k_i \hat k_j$. Now, we define
the bulk parameter $\theta=\partial_m\beta_m/3$ and  the shear parameter $\sigma=\sigma_{ij}\hat k_i\hat k_j$, 
and rewrite
  
\bea
&&2\partial_{i}\beta_{j}{\bf P}_{ij}\cos^2 \theta_p+\partial_{i}\beta_{j}{\bf N}_{ij} \sin^2 \theta_p\nonumber \\ 
&&=\sigma(3\cos^2 \theta_p-1)+2\theta
\eea
in terms of which (\ref{Z1XX}) reads

\bea
\label{Z1XXX}
\frac{d\Gamma^<_1}{d^3k}=&& -\frac{\alpha}{2\pi^2\omega}\int \frac {d^3p}{(2\pi)^3}\frac 1{2E} 
\frac {e^{\beta E}}{(e^{\beta E}-1)^2}\,\frac {p^2}{E}\nonumber\\
&&\times\left(t_\eta\sigma(3\cos^2 \theta_p-1)+2t_\zeta\theta\right)\nonumber\\
&&\times \left(-\frac{3}{f_\pi^2}(p-k)^2\text{Im} {\bf \Pi}_A \left( (p-k)^2\right) \right)\nonumber\\
\eea
after using the substitution (\ref{HYDROTIME}). 

For comparison, the equilibrium photon rates (\ref{1}) in the hadronic phase can also be calculated using the 
Wightman  function  with the result

\bea
\label{Z1XX1X}
\frac{d\Gamma^<_0}{d^3k}=&&\frac{\alpha}{\pi^2\omega}\int \frac {d^3p}{(2\pi)^3}\frac 1{2E}
\frac 1{e^{\beta E}-1}\nonumber\\
&&\times \left(\frac{3}{f_\pi^2}(p-k)^2\text{Im} {\bf \Pi}_A \left( (p-k)^2\right) \right)\nonumber\\
\eea
However (\ref{Z1XX1X}) to this order does not enforce the KMS condition,

\be
G^<(q)=\frac 2{e^{\beta q_0}+1}\,{\rm Im}\,iG^F(q)
\ee
 which reflects on the  causal
character of the emissivities. To enforce this condition  requires re-summing higher 
order contributions from the expansion in (\ref{5X8}). This is possible, and the result is~\cite{STEELE}

\bea
\label{Z1XX1}
\frac{d\Gamma_0}{d^3k}=&&\frac 1{e^{\beta \omega}+1}\frac{\alpha}{\pi^2\omega}\int \frac {d^3p}{(2\pi)^3}\frac 1{2E}
\frac 1{e^{\beta E}-1}\nonumber\\
&&\times \left(\frac{3}{f_\pi^2}(p+k)^2\text{Im} {\bf \Pi}_A \left( (p+k)^2\right) + (k\to -k)\right)\nonumber\\
\eea
The chief outcomes of this re-summation are two-fold: 1/ the appearance of an overall factor of
$1/(e^{\beta \omega}+1)$; 2/ a crossing of the spectral function in the integrand that yields the
full forward $\gamma^*\pi\rightarrow \gamma^*\pi$ Feynman amplitude. We now apply these observations to
(\ref{Z1XXX}) to obtain

\bea
\label{Z1XXXY}
\frac{d\Gamma_1}{d^3k}=&& -\frac 1{e^{\beta \omega}+1}\frac{\alpha}{2\pi^2\omega}\int \frac {d^3p}{(2\pi)^3}\frac 1{2E} 
\frac {e^{\beta E}}{(e^{\beta E}-1)^2}\,\frac {p^2}{E}\nonumber\\
&&\times\left(t_\eta\sigma(3\cos^2 \theta_p-1)+2t_\zeta\theta\right)\nonumber\\
&&\times \left(-\frac{3}{f_\pi^2}(p-k)^2\text{Im} {\bf \Pi}_A \left( (p-k)^2\right) +(k\rightarrow -k)\right)\nonumber\\
\eea
(\ref{Z1XXXY}) is our final result for
the leading viscous correction to the hadronic rate using spectral functions. 
The total viscous photon hadronic rate follows from (\ref{Z1XX1}) plus (\ref{Z1XXXY}) as

\be
\frac{d\Gamma}{d^3k}=\frac{d\Gamma_0}{d^3k}+\frac{d\Gamma_1}{d^3k}
\ee

\subsection{Vector dominance}

For a simple  estimate of the size of the  viscous corrections, we 
will use the un-summed rates $\Gamma_{0,1}^<$, and make use of
vector dominance model (VDM) to saturate ${\rm Im}\,\Pi_A$. Specifically, we  set

\be
\label{Z2XX}
{\rm Im}\,\Pi_A(s)\approx  f_A^2\,\frac{\frac{\Gamma}{2}}{(s-m_A^2)^2+\frac{\Gamma^2}{4}}
\ee
with the axial constant $f_A\approx f_\pi$. Here $m_A,\Gamma$ are the mass and width of the axial-meson.  Inserting (\ref{Z2XX}) in (\ref{Z1XXX}) yields the VDM result for the 
(un-summed) viscous photon rate

\bea
\label{Z3XX}
\frac{d\Gamma^<_1}{d^3k}\approx &&-\frac{3\alpha}{8\pi^4}\frac{f_A^2}{f_\pi^2}\,\Gamma\int \frac{p^4dp\,d\cos \theta_p}{E^2}\frac{e^{\beta E}}{(e^{\beta E}-1)^2}\nonumber \\ 
&&\times \left(\left(t_\zeta\theta-\frac{1}{2}t_\eta \sigma\right)+\frac{3}{2}t_\eta\sigma\cos^2 \theta_p\right)\nonumber\\
&&\times \frac{E-p\cos \theta_p}{(m_A^2+2E\omega-2p\omega\cos \theta_p)^2+\frac{\Gamma^2}{4}}
\eea
(\ref{Z3XX}) simplifies further as $\beta m_\pi\rightarrow 0$ (chiral limit), 

\bea
\label{Z4XX}
\frac{d\Gamma^<_1}{d^3k}\approx&& -\frac{3\alpha}{8\pi^4}\frac{f_A^2}{f_\pi^2}\,\Gamma\int_{m_{\pi}}^{\infty} E^3dE\frac{e^{\beta E}}{(e^{\beta E}-1)^2}\nonumber \\ 
&&\times \left(\left(t_\zeta\theta-\frac{1}{2}t_\eta\sigma\right)f_1(E)+\frac{3}{2}t_\eta\sigma f_2(E)\right)
\eea
where we have kept $m_\pi$ as an infrared regulator in the integration, with

\bea
\label{F12}
&&f_1(E)=\int _{-1}^1dx \frac{1-x}{(m_A^2+2E\omega(1-x))^2+\frac{\Gamma^2}{4}}\nonumber\\
&&f_2(E)=\int _{-1}^1dx \frac{x^2(1-x)}{(m_A^2+2E\omega(1-x))^2+\frac{\Gamma^2}{4}}
\eea
(\ref{Z3XX})  is seen to vanish for zero width $\Gamma$.  To leading order in $\Gamma$, (\ref{F12}) simplifies

\bea
m_A^4f_1(E)\approx f_1(x)=&&\frac 1{4x^2} \left({\rm ln(1+4x)-\frac{4x}{1+4x}}\right)\nonumber\\
m_A^4f_2(E)\approx f_2(x)=&&\frac 1{16x^4}\biggl((3+2x)(1+2x)\rm ln(1+4x)\nonumber\\
&&-8x(1+x)-\frac{4x(1+2x)^2}{1+4x}\biggr)\nonumber\\
\eea
with $x={E\omega}/{m_A^2}$. Changing the integration variable to $x$ in (\ref{Z3XX}) gives

\bea
\frac{d\Gamma^<_1}{d^3k}\approx &&-\frac{3\alpha}{16\pi^4}\frac{f_A^2m_A^4}{f_\pi^2\omega^4}\,\Gamma
 \int_{\frac{m_\pi\omega}{m_A^2}}^{\infty} dx  \frac{x^3e^{\frac{\beta m_A^2x}{\omega}}}{(e^{\frac{\beta m_A^2x}{\omega}}-1)^2}
\nonumber\\&&\times\left(2t_\zeta\theta f_1(x)+t_\eta\sigma(3f_2(x)-f_1(x)) \right)
\eea



  \begin{figure}[t]
  \begin{center}
  \includegraphics[width=8cm]{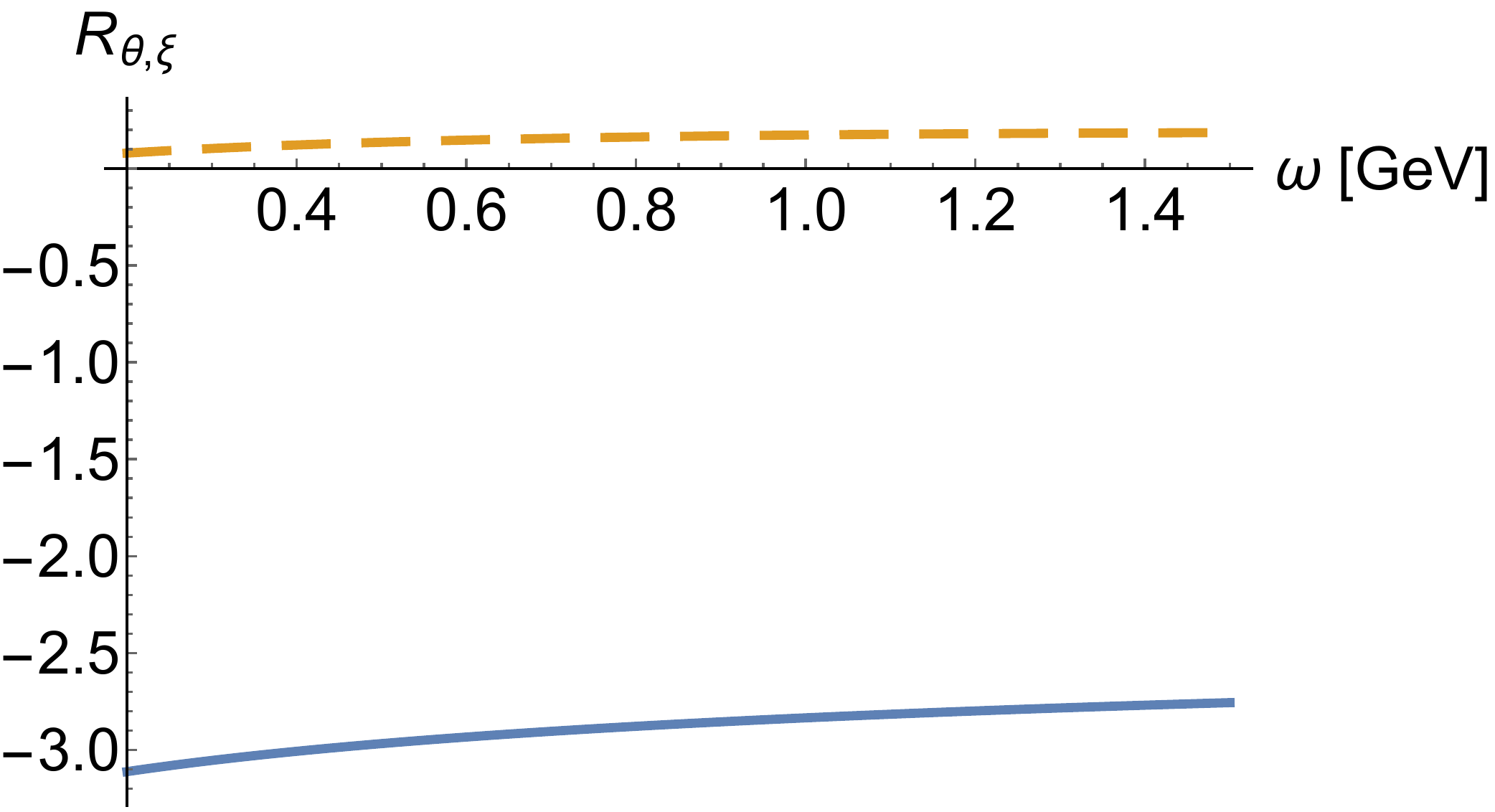}
  \caption{Ratio (\ref{Z4XX3}) for the bulk hadron contribution $R_\zeta$ blue-solid curve vs $\omega$, 
  and for the hadron shear contribution $R_\eta$  orange-dashed curve vs $\omega$  in the VDM approximation,
   for $\beta=1/m_\pi$, and 
  equal relaxation times $t_\zeta=t_\eta=\beta$ for fixed $\theta=\sigma=1$.}
    \label{fig_rate}
  \end{center}
\end{figure}

For comparison, the (un-summed) equilibrium VDM rate (\ref{Z1XX1X}) in the same approximation reads

\be
\label{Z4XX2}
\frac{d^3\tilde \Gamma^<_0}{d^3k}\approx  \frac{3\alpha }{8\pi^4}\frac{f_A^2m_A^2}{f_\pi^2\omega^3}\,\Gamma\,
\int_{\frac{m_\pi\omega}{m_A^2}}^{\infty} dx  \frac{x^2f_1(x)}{e^{\frac{\beta m_A^2x}{\omega}}-1}
\ee
where the lower bound stems  from

\be
(p+k)^2=2E\omega(1-\cos \theta_p)=m_{A}^2 
\ee
The ratio of the (un-summed) viscous rate (\ref{Z4XX}) to
the equilibrium rate (\ref{Z4XX2}) takes the simple form

\bea
\label{Z4XX3}
&&\frac{d\Gamma^<_1}{d\Gamma^<_0}\approx 
-\frac {m_A^2}{2\omega}
\left(\int_{\frac{m_\pi\omega}{m_A^2}}^{\infty} dx  \frac{x^2f_1(x)}{e^{\frac{\beta m_A^2x}{\omega}}-1}\right)^{-1}\nonumber \\
&&\times\biggl( 2t_{\zeta}  \theta \int_{\frac{m_\pi\omega}{m_A^2}}^{\infty} dx  \frac{x^3e^{\frac{\beta m_A^2x}{\omega}}}{(e^{\frac{\beta m_A^2x}{\omega}}-1)^2} f_1(x)\biggr.\nonumber \\ 
&&\biggl.\,\,\,\,\,\,+t_{\eta}\sigma\int_{\frac{m_\pi\omega}{m_A^2}}^{\infty} dx  \frac{x^3e^{\frac{\beta m_A^2x}{\omega}}}{(e^{\frac{\beta m_A^2x}{\omega}}-1)^2}(3f_2(x)-f_1(x)) \biggr)\nonumber\\
\eea

In Fig.~\ref{fig_rate} we show the bulk contribution $R_\zeta$ in (\ref{Z4XX3}) as the solid-blue curve, and the shear contribution 
$R_\eta$ in (\ref{Z4XX3}) as the orange-dashed over a range of frequencies $\omega$ in GeV for a temperature 
$1/\beta=m_\pi$. We have set the shear and bulk factors to $\theta=\sigma=1$, and fixed the relaxation times to
$t_\eta=t_\zeta=\beta$.
The smallness of the shear contribution stems from the near cancellation of the $3f_2-f_1$ in the integrand of (\ref{Z4XX3}).
The bulk contribution dwarfs the shear contribution in the VDM approximation for about all frequencies. The bulk contribution
is also opposite in sign to the shear contribution in leading order. 
At currently available collider energies, a typical AA collision triggers a hadronic fluid with a size $L\approx 10$ fm. 
For temperatures $T\approx 200$ MeV that results in fluid gradients of the size $\theta\approx \sigma\approx 1/TL\approx 1/10$.
When combined with the result shown in Fig.~\ref{fig_rate}, this estimate shows that the bulk viscosity correction to the hadronic rate is 
about 30\% across all frequencies, while the shear viscosity correction is negligible. 
Overall, the bulk hydrodynamical correction appears sizable even in the late stage of the hadronic evolution with smal gradients 
in the form of a small $\theta=\partial_i\beta_i/3$. 
These observations deserve to be further checked in current hydrodynamical assessments of the electromagnetic emissivities
and without the VDM approximation through the use of the full axial spectral weight.

\subsection{Viscous dilepton rate}

The previous results, extend to the dilepton rates as well if we recall that for dilepton emissivities (\ref{1}) need to
be changed to

\be
\label{DI1}
\frac {d\mathbb R_0}{d^4q}=-\frac {\alpha^2{\mathbb B}\,g^{\mu\nu}}{6\pi^3q^2}\left<\mathbb G_{\mu\nu}^{<}(q)\right>_0
\ee
with  the leptonic factor

\be
\mathbb B=\left(1+\frac{2m_l^2}{q^2}\right)\left(1-\frac{4m_l^2}{q^2}\right)^{\frac 12}
\ee
with the treshold $q^2>4m_l^2$ and typically $l=e,\mu$. The equilibrium contributions to (\ref{DI1})
in the hadronic phase have been discussed in details using spectral functions 
in~\cite{DILEPTON}  and hadronic processes in~\cite{HADRON}. From the spectral functions
analysis the result is~\cite{DILEPTON}

\bea
\label{5X166XX}
\frac{d\mathbb R_0}{d^4q}&&=-\frac{\alpha^2}{6\pi^3}\frac{\mathbb B}{q^2}\,
\frac {2}{e^{\beta\omega}+1}\,\nonumber\\
&&\times \biggl(-3q^2{\rm Im}\Pi_V(q^2)+\int \frac{d^3p}{(2\pi)^3}\frac 1{2E}\frac 1{e^{\beta E}-1}\nonumber\\
&& \times\biggl(\frac{12}{f_\pi^2}q^2\text{Im} {\bf \Pi}_V(q^2)\nonumber\\
&&-\frac{6}{f_\pi^2}(p-q)^2\text{Im} {\bf \Pi}_A \left( (p-q)^2+q\rightarrow -q\right)\nonumber\\
&&+\frac{8}{f_\pi^2}\left( (p\cdot q)^2-m_\pi^2 q^2\right) \text{Im} {\bf \Pi}_V(q^2)\times\text{Re} \Delta_R(p-q)\nonumber\\
&&+q\rightarrow -q \biggr)\biggr)
\eea

The non-equilibrium viscous correction in the hadronic phase, follows a similar reasoning as that
given for the photons. A rerun of the
preceding reasoning shows that $G^<_{ij, 0\pi}$ also vanishes in this case. However, 
$G^<_{1\pi}$  does not and the result is

\bea
\label{5X166}
&&(d\pi)_{ij}G^<_{1\pi}(q,k)=\frac {d^3k}{(2\pi)^3}\frac{t_H}{2E}\left(\frac{k_ik_j}{E_k}\right)n_B(1+n_B)\nonumber\\&&\times\biggl(
\frac{12}{f_\pi^2}q^2\text{Im} {\bf \Pi}_V(q^2)\nonumber\\
&&-\frac{6}{f_\pi^2}(k-q)^2\text{Im} {\bf \Pi}_A \left( (k-q)^2\right) \nonumber\\
&&+\frac{8}{f_\pi^2}\left( (k\cdot q)^2-m_\pi^2 q^2\right) \text{Im} {\bf \Pi}_V(q^2)\times\text{Re} \Delta_R(k-q)\biggr)\nonumber\\
\eea
Here $\Pi_V$ is the VV correlation of the 
vector current in the vacuum, and $\Delta_R$ is the retarded pion propagator~\cite{STEELE}. The spectral form 
of $\Pi_V$ follows from $e^+e^-$ annihilation.  The last bracket in (\ref{5X166}) is only the {\it crossed }
$\gamma^*\pi\rightarrow \gamma^*\pi$ scattering amplitude ${\cal T}_{\gamma^*\pi}$,
which is seen to reduce to (\ref{5X16}) at the photon point or $q^2=0$.  In terms of (\ref{5X166}) the
re-summed viscous corrections to the dilepton emissivities in a hadronic fluid take the following final form

\bea
\label{5X166X}
\frac{d\mathbb R_1}{d^4q}&&=
-\frac{4\alpha^2}{3\pi^2}\frac{\mathbb B}{q^2}\,\frac 1{e^{\beta\omega}+1}\,
\int \frac {p^4dp\,d{\rm cos}\theta_p}{(2\pi)^3}\frac 1{2E} \frac {e^{\beta E}}{(e^{\beta E}-1)^2}\nonumber\\
&&\times \left(\left(t_\zeta\theta-\frac{1}{2}t_\eta \sigma\right)+\frac{3}{2}t_\eta\sigma\cos^2 \theta_p\right)\nonumber\\
&& \times\biggl(\frac{12}{f_\pi^2}q^2\text{Im} {\bf \Pi}_V(q^2)\nonumber\\
&&-\frac{6}{f_\pi^2}(p-q)^2\text{Im} {\bf \Pi}_A \left( (p-q)^2+q\rightarrow -q\right)\nonumber\\
&&+\frac{8}{f_\pi^2}\left( (p\cdot q)^2-m_\pi^2 q^2\right) \text{Im} {\bf \Pi}_V(q^2)\times\text{Re} \Delta_R(p-q)\nonumber\\
&&+q\rightarrow -q \biggr)
\eea
The total viscous hadronic rate for dilepton emission is (\ref{5X166XX}) plus (\ref{5X166X}) 

\be
\frac{d\mathbb R}{d^4q}=\frac{d\mathbb R_0}{d^4q}+\frac{d\mathbb R_1}{d^4q}
\ee

\section{Partonic phase}

At high temperature the fluid is that of strongly coupled partonic-like constituents (sQGP). We will treat it  in leading order
as made of partonic constituents in the presence of soft gluonic fields. The soft corrections will be estimated as operator
insertions in leading dimensions as in~\cite{OPE}.  A similar proposal using soft insertions for the electromagnetic emissivities
was also suggested in~\cite{ROB}. With this in mind, and 
for the generic process $[p_i]\rightarrow [q_f]+\gamma$, the unordered Wightman function reads~\cite{LARRY},

 \bea
 \label{GENERAL}
 &&-G^{\mu<}_{\mu}(q)=\nonumber \\ 
 &&\int \prod _{i}\frac{d^3p^{\rm in}_{i}}{(2\pi)^32E_{i}^{\rm in}}n(E_{i}^{\rm in})\prod _{j}\frac{d^3q^{\rm out}_{j}}{(2\pi)^32E_{j}^{\rm out}}(1\pm n(E_{j}^{\rm out}))\nonumber \\ 
&&\times (2\pi)^4\delta\left(\sum _i{p_i}-\sum_j {q_j}-{q}\right)\, |{\bf M}_{i\rightarrow f+\gamma}|^2
 \eea
 The
effects of the viscous corrections amount to additional contributions to the initial and final distribution
functions. We now detail them for both dilepton and photon emissions.

\subsection{Dileptons}

We  now seek to organize the photon emissivities in the non-perturbative partonic phase as follows

\be
\frac{d^3\mathbb R}{d^4q}=\frac{d^3\mathbb R^{Tp}_0}{d^4q}+\frac{d^3\mathbb R^{Vp}_1}{d^4q}+\frac{d^3\mathbb R^{Tn}_0}{d^4q}+\frac{d^3\mathbb R^{Vn}_1}{d^4q}
\ee
with the first contribution $\mathbb R^{Tp}_0$ as the thermal perturbative rate, 
the second contribution $\mathbb R^{Vp}_1$ as the viscous perturbative correction,
the third contribution $\mathbb R^{Tn}_0$ as the thermal and non-perturbative correction of leading mass dimension in the external fields,  and finally the fourth contribution  $\mathbb R^{Vp}_1$ as the viscous non-perturbative contribution in leading mass dimension in the external fields.  We now proceed to evaluate each of these contributions sequentially as (\ref{P2}), (\ref{P2X1}), (\ref{P4}) and (\ref{P2X2}) to be detailed below.

 \subsubsection{Thermal perturbative contribution}

\begin{figure}[b]
  \begin{center}
  \includegraphics[width=8cm]{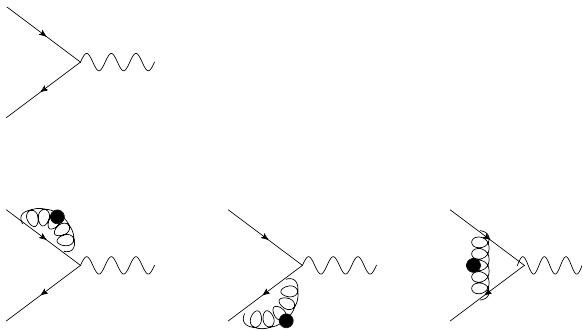}
  \caption{Thermal dilepton emission including the leading perturbative term (top) and  the 
  leading soft and non-perturbative corrections (bottom). The blob refers to gluon insertions
   of leading dimensions $(gA_4)^2, (gE)^2, (gB)^2$. }
  \label{fig_dilepton}
  \end{center}
\end{figure}

In leading order, the perturbative dilepton emissivity corresponds to an in-state with a single $q\bar q$
as illustrated in Fig.~\ref{fig_dilepton},
and its contribution to (\ref{GENERAL}) is (omitting all charge factors)

 \bea
 \label{DEE1}
 &&-G^{\mu<}_{\mu}(q)=\frac{q^2}{\pi|q|}\int_{\frac{q_0-|q|}{2}}^{\frac{q_0+|q|}{2}}dkf_{\mu}(k)f_{-\mu}(q_0-k)\nonumber \\ 
 &&=n_B(q_0)\,\frac{q^2}{\pi |q|}\int_{q^-}^{q^+} (1-f_{\mu}(k)-f_{-\mu}(q_0-k))\nonumber \\
 && =n_B(q_0)\,\frac{q^2}{\pi }n_B(q_0)\left(-1+\frac{1}{\beta|q|}\rm ln\left(\frac{n_{\mu}^-n_{-\mu}^-}{n_{\mu}^+n_{-\mu}^+}\right)\right)
 \eea
 we have defined the Fermi distributions at finite {\it chemical potential} $\mu$ as

\be
f_{\pm \mu}(q)=\frac 1{e^{\beta({q_0\mp \mu})}+1}
\ee
and their associated shifted distributions

\be
 \label{P3}
 n_{\pm \mu}^\pm =\frac 1{e^{\beta(q_0\pm |q|)/2\mp \beta\mu}+1}
 \ee
 The emergence of the Bose distribution $n_B=1/(e^{\beta q_0}-1)$ in (\ref{DEE1}) reflects on the
KMS condition
 
 \be
 \label{RETX}
 G^{\mu<}_{\mu}(q)=2n_B(q_0)\,{\rm Im}\,iG^{\mu R}_{\mu}(q)
 \ee
 at finite temperature and chemical potential $\mu$ in leading order.
The  finite chemical potential will be traded below for a complex chemical potential for 
a fixed color species and identified with the insertion of a soft $A_4$ contribution in
the strongly coupled QGP~\cite{OPE,ROB}.  For $\mu=0$, (\ref{DEE1})  
when inserted in the general formula for dilepton emission (\ref{DI1}) and upon 
restoring the color-flavor factor for partons $N_c\hat e_f^2/2$, yield
the leading partonic dilepton rate

 \bea 
 \label{P2}
 \frac{d\mathbb R^{T}_0}{d^4q}&&=\frac {-\alpha^2\mathbb B}{3\pi^3}\frac 1{e^{\beta q_0}-1}\nonumber\\
 &&\times \left(\frac 1{4\pi}\,d_F\sum_f \hat e_f^2\right)
 \left(1+\frac 2{\beta |q|}\,{\rm ln}\left(\frac{n^+}{n^-}\right)\right)
 \eea
with $d_F=N_c$ the color dimension of the quark representation,
and $n^\pm \equiv n^\pm_{\pm 0}$. Here $e\hat e_f$ is the electromagnetic charge of a quark of flavor $f$.
 In this order, the emission is isotropic.

 \begin{figure}[t]
  \begin{center}
  \includegraphics[width=8cm]{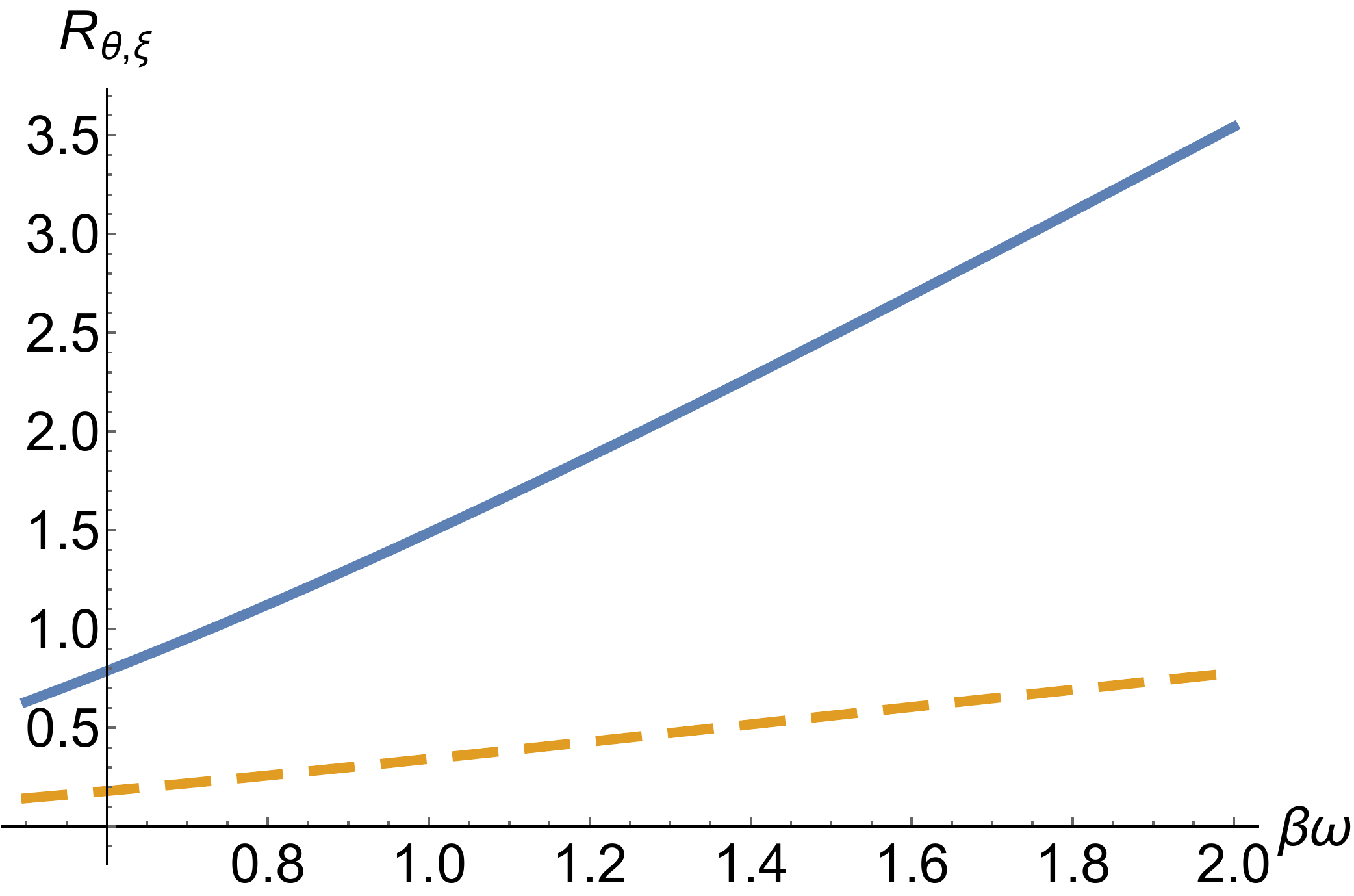}
  \caption{Ratio $d{\mathbb R_1^{Vp}}/d{\mathbb R_0^{Tp}}$  for the thermal perturbative dilepton
  bulk contribution $R_\zeta$ blue-solid curve vs $\beta\omega$, 
  and for the thermal perturbative shear dilepton contribution $R_\eta$  orange-dashed curve vs $\beta\omega$
  and $q=(2\omega, \omega)$. We have set the 
   relaxation times $t_\zeta=t_\eta=\beta$ and  fixed $\theta=\sigma=1$.}
    \label{fig_rate_parton}
  \end{center}
\end{figure}

 \subsubsection{Viscous perturbative contribution}

The viscous corrections to the perturbative quark and gluon processes, follow exactly
along the general arguments we presented earlier in sections II and III. Specifically, for the
Fermionic Wightman functions

\bea
&&G_{\alpha \beta}^{>}(x)=+\left<\psi_{\alpha}(x)\bar \psi_{\beta}(0)\right>\nonumber\\
&&G_{\alpha \beta}^{<}(x)=-\left<\bar \psi_{\beta}(0) \psi_{\alpha}(x)\right>
\eea
the $T_{ij}$ insertions  amounts to additional contributions, and  in leading order we have

\bea
&&G^{<}(k)=-\frac{\pi \slashed{k}}{E}n(k_0)(\delta^--\delta^+)\nonumber \\ 
&&-t_H\partial_{i}\beta_{j}\frac{k^{i}k^{j}}E\frac{\pi\slashed{k}}{E}n(k_0)(1-n(k_0))(\delta^-+\delta^+)\nonumber\\
&&G^{>}(k)=+\frac{\pi \slashed{k}}{E}(1-n)(\delta^--\delta^+)\nonumber \\ 
&&-t_H\partial_{i}\beta_{j}\frac{k^{i}k^{j}}E\frac{\pi\slashed{k}}{E}n(1-n)(\delta^-+\delta^+)
\eea
with $\delta^\pm\equiv\delta(k_0\mp  k)$. In the real-time or double-line formalism,
the total emission rate follows from the 12 Wightman function, where the effects 
of the $T_{ij}$ insertions amount to modifying the {\bf in-state} population by

 \be
 \label{001}
 n\rightarrow n+t_H\frac{k_ik_j}E\partial_{i}\beta_j\,n(1\pm n)
 \ee
and the {\bf out-state} population by

 \be 
 \label{002}
 n\rightarrow 1-n-t_H\frac{k_ik_j}E\partial_{i}\beta_j\,n(1\pm n)
 \ee
 With this in mind, the viscouss corrections to the leading order dilepton emission 
 at finite chemical potential (\ref{DEE1}) is

 \bea
 \label{YYTH}
&&+t_H \partial_{i}\beta_j\frac{q^2}{\pi|q|}\int_{\frac{q_0-|q|}{2}}^{\frac{q_0+|q|}{2}}dkf_{\mu}(1-f_{\mu})f_{-\mu}(q_0-k)\frac{k_ik_j}{k}\nonumber \\ 
&&+t_H \partial_{i}\beta_j\frac{q^2}{\pi|q|}\int_{\frac{q_0-|q|}{2}}^{\frac{q_0+|q|}{2}}dkf_{-\mu}(1-f_{-\mu})f_{\mu}(q_0-k)\frac{k_ik_j}{k}
\nonumber\\
 \eea
 which can be re-organized as follows

\bea
\label{DEE2}
&&t_H \partial_{i}\beta_j\frac{q^2}{\pi|q|}\nonumber\\
&&\times\biggl(\frac{ \delta_{ij}-\hat q_i\hat q_j}{2}\int_{\frac{q_0-|q|}{2}}^{\frac{q_0+|q|}{2}} dk kf_{\mu}(1-f_{\mu})f_{-\mu}(q_0-k)\biggr.\nonumber \\ 
&&\biggl.+\frac{3\hat q_i\hat q_j-\delta_{ij}}{2} \int_{\frac{q_0-|q|}{2}}^{\frac{q_0+|q|}{2}} dk kf_{\mu}(1-f_{\mu})f_{-\mu}(q_0-k)\biggr.\nonumber\\
&&\biggl.\times \left(\frac{q_0}{|q|}-2\frac{q_+q_-}{|q|k}\right)^2+\mu\rightarrow -\mu\biggr)
\eea
The $\mu=0$ contribution in (\ref{DEE2}) yields the viscous perturbative contribution to the dilepton rate 
(\ref{P2}).  More specifically, we have

 \bea 
 \label{P2X1}
 \frac{d\mathbb R^{Vp}_1}{d^4q}&&=\frac {\alpha^2\mathbb B}{3\pi^3}\frac 1{e^{\beta q_0}-1}\nonumber\\
  &&\times \left(\frac 1{4\pi}\,d_F\sum_f \hat e_f^2\right)\, \frac 2{|q|}\nonumber\\
&&\times\biggl(\frac{ 2t_\zeta\theta-t_\eta\sigma}{2}\int_{\frac{q_0-|q|}{2}}^{\frac{q_0+|q|}{2}} dk k(1-f-\tilde f)(1-f)\biggr.\nonumber \\ 
&&\biggl.+\frac{3t_\eta\sigma}{2} \int_{\frac{q_0-|q|}{2}}^{\frac{q_0+|q|}{2}} dk k (1-f-\tilde f)(1-f)\biggr.\nonumber\\
&&\biggl.\times \left(\frac{q_0}{|q|}-2\frac{q_+q_-}{|q|k}\right)^2\biggr)
\nonumber\\
 \eea
where we have defined the fermionic distributions $f\equiv f_0(k)$ and $\tilde f\equiv f_0(q_0-k)$ for $\mu=0$. 

In Fig.~\ref{fig_rate_parton} we show the ratio $d{\mathbb R_1^{Vp}}/d{\mathbb R_0^{Tp}}$ of the thermal
viscous  contribution (\ref{P2X1}) to the free thermal contribution (\ref{P2}),
for dilepton emission at $q=(2\omega, \omega)$ as a function of $\beta\omega$,
after setting $t_\zeta=t_\eta=\beta$ and $\theta=\sigma=1$. The orange-dashed line is the shear ratio,
while the blue-solid line is the bulk ratio. Again, the bulk contribution is larger than the shear contribution and both 
are positive and increasing with $\beta\omega$.

 \subsubsection{Thermal non-perturbative contribution}

The  partonic phase near the transition
temperature still carries soft gluons~\cite{OPE,ROB}.
Their effects  is to modify  both the thermal and viscous rates. 
A way to assess these non-perturbative effects is to organize these
modifications as power corrections through gluonic 
operators insertions of increasing dimension $\Delta=2,4$ 
 in the  $JJ$ correlation function. In Fig.~\ref{fig_dilepton} we illustrate the leading soft gluonic insertions
 on the dilepton emissivities.
Typically, these contributions are  of the form $(gA_4)^2, (gE)^2, (gB)^2, ...$ and of order $(g^2T)^{\Delta}$.

Since a constant $gA_4$ acts as an imaginary colored  chemical potential on the quark line, 
the leading operator insertion $(gA_4)^2$  is readily obtained from the quadratic $\mu$-contribution steming
from the  fermionic propagator at finite chemical potential, with the identification

\be
\label{S1}
\mu^2\rightarrow -\left<(gA_4)^2\right>
\ee
A proof of this is given in the Appendix A using the background field method.
With this in mind, the quadratic contribution steming from (\ref{DEE1}) is

 \bea
 &&\rm ln\left(\frac{n_{\mu}^-n_{-\mu}^-}{n_{\mu}^+n_{-\mu}^+}\right)=+2\,\rm ln\left(\frac{n^-}{n^+}\right)\nonumber\\
&&+(\beta\mu)^2(n^{+}(1-n^{+})-n^{-}(1-n^{-})) +{\cal O}(\mu^4)
 \eea
which corrects the perturbative dilepton rate (\ref{P2})  by the non-perturbative contribution

 \bea 
 \label{P4}
&&-\frac {\alpha^2\mathbb B}{3\pi^3}\frac 1{e^{\beta q_0}-1}
\left(\frac 1{4\pi}\,d_F\sum_f \hat e_f^2\right)\left<(gA_4)^2\right>\nonumber\\
&&\times \left(\frac{\beta}{|q|}\right)\left(n^+(1-n^+)-n^-(1-n^-)\right)
 \eea
The effects of $(gE)^2,(gB)^2$ can be calculated by general arguments using the background field method~\cite{OPE},
as briefly recalled in the Appendix.
The net result can be understood using the following simple substitution

\bea
\label{S2}
\left<(gA_4)^2\right>\rightarrow \left<(gA_4)^2\right> - \frac 1{6q^2}\left<(gE)^2\right>
+ \frac 1{3q^2}\left<(gB)^2\right>
\eea
a proof of which is given in  Appendix A.
The substitution  can be understood as $(gqA_4)\sim gE\sim gB$. The factor of $\frac 13$ is from averaging over
the vector orientations. The extra $-\frac 12$ in front of the electric contribution is due to the use of a fixed
thermal frame and the fact that $(gE)^2\sim -(gB)^2$ in Euclidean space. 
Hence the final non-perturbative corrections to the dilepton rate (\ref{P2}) in leading
operator insertions

 \bea 
 \label{P4}
 \frac{d{\mathbb R}^{Tn}_0}{d^4q}&&=\frac {\alpha^2\mathbb B}{3\pi^3}\frac 1{e^{\beta q_0}-1}
\left(\frac 1{4\pi}\,d_F\sum_f \hat e_f^2\right)\nonumber\\
&&\times\left(-\left<(gA_4)^2\right> + \frac 1{6q^2}\left<(gE)^2\right>
- \frac 1{3q^2}\left<(gB)^2\right>\right)\nonumber\\
&&\times \left(\frac{\beta}{|q|}\right)\left(n_+(1-n_+)-n_-(1-n_-)\right)\biggr)
 \eea
 in agreement with the result in~\cite{OPE}. The typical values  of the  soft condensate insertions in
 (\ref{P4}) are discussed in~\cite{OPE,Lee14}.

 \subsubsection{Viscous non-perturbative contribution}

The viscous and non-perturbative corrections to (\ref{P2X1}) can be obtained using the same reasoning
developed above for the non-perturbative thermal corrections. For that, we expand the general result (\ref{DEE2}) 
to quadratic order in $\mu$, by expanding the fermionic occupation number

\be
f_{\mu}=f+ \beta\mu f(1-f)+\frac  12{(\beta\mu)^2}\,f(1-f)(1-2f) +{\cal O}(\mu^3)\nonumber\\
\ee
Now we use the identity 

\be
&&\int f_{\mu}(1-f_{\mu})f_{-\mu}(q_0-k) \,[...]=\nonumber \\ 
&&n_{B}(q_0)\int (1-f_{\mu}-f_{-\mu}(q_0-k))(1-f_{\mu})\,[...]
\ee
and expand the integrand   in $\mu$. The quadratic contribution reads

\bea
&&\biggl((1-f_{\mu}-\tilde f_{-\mu})(1-f_{\mu})+(\mu \rightarrow -\mu)\biggr)_{\mu^2}=\nonumber\\
&&+2f(1-f)^2(3f-1)+f(1-f)(1-2f)(\tilde f-1)\nonumber \\ 
&&+\tilde f(1-\tilde f)(1-2\tilde f)(f-1)-2f(1-f)\tilde f(1-\tilde f)\nonumber\\
\eea
With the help of the identity

\bea
&&2f(1-f)^2(3f-1)=\nonumber \\ 
&&(1+n_{B}(q_0))(1-\tilde f-f)(f(1-2f)+\tilde f(1-2\tilde f))\nonumber \\ 
&&-2n_{B}(q_0)(1+n_{B}(q_0)(1-f-\tilde f)^2
\eea
we have finally for the $\mu^2$ correction to the viscous corrections to (\ref{YYTH}) as

\bea
\label{000}
&&(\beta \mu)^2 t_H \partial_{i}\beta_j\frac{q^2}{\pi|q|}\times  \nonumber \\
&& \biggl(\frac{ \delta_{ij}-\hat q_i\hat q_j}{2}(n_B F_1+n_B(1+n_B)F_2+n_B^2(1+n_B)F_3)\biggr.\nonumber \\
&&\biggl.+\frac{3\hat q_i\hat q_j-\delta_{ij}}{2}(n_B \tilde F_1+n_B(1+n_B)\tilde F_2+n_B^2(1+n_B)\tilde F_3)\biggr) \nonumber\\
\eea
where we have defined

\bea
F_1=&&+2\int dk k f(1-f)^2(3f-1)\nonumber\\
\tilde F_1=&&+2\int dk k f(1-f)^2(3f-1) \left(\frac{q_0}{|q|}-\frac{2q_+q_-}{|q|k}\right)^2 \nonumber\\
F_2=&&-\int dk k(1-\tilde f -f)(f(1-2f)+\tilde  f(1-2\tilde f)) \nonumber\\
\tilde F_2=&& -\int dk k(1-\tilde f -f)(f(1-2f)+\tilde  f(1-2\tilde f)) \nonumber \\ 
&&\times \left(\frac{q_0}{|q|}-\frac{2q_+q_-}{|q|k}\right)^2 \nonumber\\
F_3=&&-2\int dk k(1-\tilde f -f)^2\ \nonumber\\
\tilde F_3=&&-2\int dk k(1-\tilde f -f)^2\left(\frac{q_0}{|q|}-\frac{2q_+q_-}{|q|k}\right)^2
\eea
Using the operator substitutions (\ref{S1}-\ref{S2}) for $\mu^2$ in (\ref{000}) lead to the non-perturbative 
corrections to the viscous dilepton emission rate (\ref{P2X1}) in the form  $(n_B\equiv n_B(q_0))$

 \bea 
 \label{P2X2}
&& \frac{d\mathbb R^{Vn}_1}{d^4q}=\frac {\alpha^2\mathbb B}{3\pi^3q^2}\frac {\beta^2}{e^{\beta q_0}-1}\nonumber\\
  &&\times \left(\frac 1{4\pi}\,d_F\sum_f \hat e_f^2\right)\, \frac {q^2}{\pi |q|}\nonumber\\
&&\times \left(-\left<(gA_4)^2\right> + \frac 1{6q^2}\left<(gE)^2\right>
- \frac 1{3q^2}\left<(gB)^2\right>\right)\nonumber\\
&&\biggl( \frac{ 2t_\zeta\theta-t_\eta\sigma}{2}(n_B F_1+n_B(1+n_B)F_2+n_B^2(1+n_B)F_3)\biggr.\nonumber \\
&&\biggl.+\frac{3t_\eta\sigma}{2}(n_B \tilde F_1+n_B(1+n_B)\tilde F_2+n_B^2(1+n_B)\tilde F_3)\biggr) \nonumber\\
\eea

\begin{figure}[b]
  \begin{center}
  \includegraphics[width=8cm]{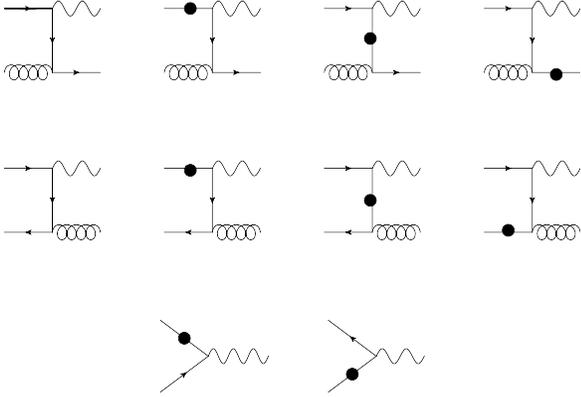}
  \caption{Thermal photon emission including the leading perturbative term   and  the 
  leading soft and non-perturbative corrections. The blob refers to gluon insertions
   of leading dimensions $(gA_4)^2, (gE)^2, (gB)^2$. Soft vertex insertions as in Fig.\ref{fig_dilepton}
   are also
   included but not shown. The last two contributions are not
   allowed without the soft insertions. }
  \label{fig_photon}
  \end{center}
\end{figure}

\subsection{Photons}

Following the dilepton analysis, we now seek to organize the photon emissivities in the non-perturbative partonic phase as follows

\be
\frac{d^3\Gamma}{d^3k}=\frac{d^3\Gamma^{Tp}_0}{d^3k}+\frac{d^3\Gamma^{Vp}_1}{d^3k}+\frac{d^3\Gamma^{Tn}_0}{d^3k}+\frac{d^3\Gamma^{Vn}_1}{d^3k}
\ee
with the first contribution $\Gamma^{Tp}_0$ as the thermal perturbative rate, 
the second contribution $\Gamma^{Vp}_1$ as the viscous perturbative correction,
the third contribution $\Gamma^{Tn}_0$ as the thermal and non-perturbative correction of leading mass dimension in the external fields,  and finally the fourth contribution  $\Gamma^{Vp}_1$ as the viscous non-perturbative contribution in leading mass dimension in the external fields.  We now proceed to evaluate each of these contributions sequentially as (\ref{P1}), (\ref{FCPXX}), (\ref{P4XY}) and 
(\ref{VNX}) to be detailed below.

\subsubsection{General}

The photon analysis is more involved since the in-state with $q\bar q$ is kinematically not allowed. 
The partonic photon emission proceeds through:
1/ the Compton channel, with $qg\rightarrow \gamma q$ or  $\bar qg\rightarrow \gamma \bar q$;
2/ the pair annihilation channel with $\bar q q\rightarrow g\gamma$, as illustrated in Fig.~\ref{fig_photon}.
 Specifically, we have

\bea
\label{VCV}
&&G^{<}_{\rm compton}=\frac{{\bf N}}{8(2\pi)^4E}\int ds dt|M_c(s,t)|^2\nonumber \\ 
&&\times \int dE_1dE_2f_{\mu}(E_1)f_g(E_2)(1-f_{\mu}(E_1+E_2-E))\nonumber \\ 
&&\times \frac{\theta(E_1+E_2-E)}{\sqrt{aE_1^2+bE_1+c}}+\mu \rightarrow -\mu
\eea
with $E=\omega=k$ throughout, and

 \be
 \label{VPV}
&& G^{<}_{\rm pair}=\frac{{\bf N}}{8(2\pi)^4E}\int ds dt|M_p(s,t)|^2\nonumber \\ 
&&\times \int dE_1dE_2f_{\mu}(E_1)f_{-\mu}(E_2)(1+f_{g}(E_1+E_2-E))\nonumber \\
&& \times \frac{\theta(E_1+E_2-E)}{\sqrt{aE_1^2+bE_1+c}}
 \ee
The leading perturbative contributions to the squared matrix elements are

\bea
&&\frac{|M_{c}(s,t,u)|^2}{16\pi^2}=-8\alpha \alpha_s\frac{u^2+s^2}{us}\nonumber\\
&&\frac{|M_{p}(s,t,u)|^2}{16\pi^2}=+8\alpha \alpha_s\frac{u^2+t^2}{ut}
\eea
and the color-flavor factor is

\be
{\bf N}=\frac{N_c^2-1}{2N_c}N_c\sum_{f}\hat e_f^2\equiv C_Fd_F\sum_{f}\hat e_f^2
\ee
The Mandelstam variables and the kinematical parameters $a,b,c$ are collectively defined as

 \bea
&&s=(p_1+p_2)^2\nonumber\\
&&t=(p-p_1)^2\nonumber\\
 &&a=-(s+t)^2\nonumber\\
&& b=2(s+t)(Es-E_2t)\nonumber\\
&& c=st(s+t)-(Es+E_2t)^2
 \eea
with $s+t+u=0$. The range of the integrations  are $s\geq 0$ and $-s\leq t\leq 0$. However, the $s$-integration is 
infrared sensitive, so the integration range will be modified to $s\geq m_T^2$ with the squared thermal quark mass 
$m_T^2=\pi\alpha_s C_F/\beta^2$ as a regulator.  These results are
in agreement with those first reported in~\cite{NAIVE}.

\subsubsection{Thermal perturbative contribution}

In this section we will detail the approximations in the reduction of (\ref{VCV}-\ref{VPV}) in leading order,
as they will be used for the viscous contributions as well.
Following~\cite{NAIVE} we can unwind the integrations through the Boltzmann approximation

\be
\label{APPX}
f_0(E_1)f_g(E_2)\sim e^{-\beta(E_1+E_2)}\sim e^{-\beta E}
\ee
in terms of which the integrand is typically of the form

\be
\label{INT}
\int dE_1dE_2 f(E_1+E_2-E)\frac{\theta(E_1+E_2-E)}{\sqrt{aE_1^2+bE_1+c}}
\ee 
After the change of variables $E_1=x^{\prime}+y, E_2=x^{\prime}-y$, this integral simplifies

 \be
 2\int dx^\prime \int dy\frac{f(2x^{\prime}-E)}{\sqrt{s^2(y-y_2)(y_1-y)}} 
 \ee 
with the integration over y giving just  $\frac{\pi }{s}$. From the constraint

\be
2(E_1+E_2-E)E(1-\cos \theta_{34})=s
\ee
we find that $2x^{\prime}-E=2x\ge \frac{s}{8E}$, and (\ref{INT}) gives

\be
\frac{2\pi}{s} \int_{\frac{s}{8E}}^{\infty}dx\,f(2x)  
\ee
For either distributions $f=\frac{1}{e^{\beta x}\pm 1}$ we obtain

\be
\frac{\pi }{\beta s}\rm ln(1\pm e^{-\frac{\beta s}{4E}})^{\pm 1}
\ee
The $t$-integrations can be carried explicitly with the results

\be
&&\int \frac{dt}s {|M_c(s,t)|^2}\nonumber \\ 
&&=128\pi^2\alpha \alpha_s\left({\rm ln}\left(\frac{s-m_T^2}{m_T^2}\right)+\frac{1}{2}\left(1-\frac{2m_T^2}{s}\right)\right)\nonumber\\
&&\int \frac{dt}s {|M_p(s,t)|^2}\nonumber \\ 
&&=256\pi^2\alpha \alpha_s\left({\rm ln}\left(\frac{s-m_T^2}{m_T^2}\right)-\left(1-\frac{2m_T^2}{s}\right)\right)
\ee
Again, the infrared cutoff satisfies $2m_T^2\le s$ ,$-s+m_T^2\le t\le -m_T^2$.
With the above in mind, the leading equilibrium photon emission from a perturbative QCD plasma associated to 
the Compton  $q g\rightarrow q\gamma$ and pair creation $q\bar q\rightarrow \gamma g$ processes, is~\cite{NAIVE}

\bea
\label{P1}
\frac{d\Gamma^{T}_0}{d^3k}&&=\frac {\alpha\alpha_s}{\pi}\frac 1{E\beta^2}\frac{1}{e^{\beta E}-1}\nonumber\\
&&\times \left(\frac 1{4\pi}\,C_Fd_F\sum_f\hat e_f^2\right)\,\left(\frac 12 {\rm ln}\left(\frac {4E }{\beta m_T^2}\right)+{\rm }
{\bf C}\right)\nonumber\\
\eea
 Here ${\bf C}$ is a constant. The emission rate to this order is isotropic.  
In (\ref{P1}) the overall substitution 

\be
\label{KMS1}
e^{-\beta E}\rightarrow \frac 1{e^{\beta E}-1}
\ee
was made to recover the causal pre-factor required by the KMS condition for the retarded process as in (\ref{RETX}).

Finally, we remark that the perturbative photon  rate (\ref{P1}) 
receives  additional  perturbative corrections to the same order  in $\alpha\alpha_s$ 
through collinear Bremsstralung~\cite{AMY}. This effect
 and its resummation will not be discussed here.   Instead, we will focus on the potentially soft gluonic corrections that are
also important near the transition temperature as we now detail.

\subsubsection{Viscous perturbative  contributions}

\begin{figure}[t]
  \begin{center}
  \includegraphics[width=8cm]{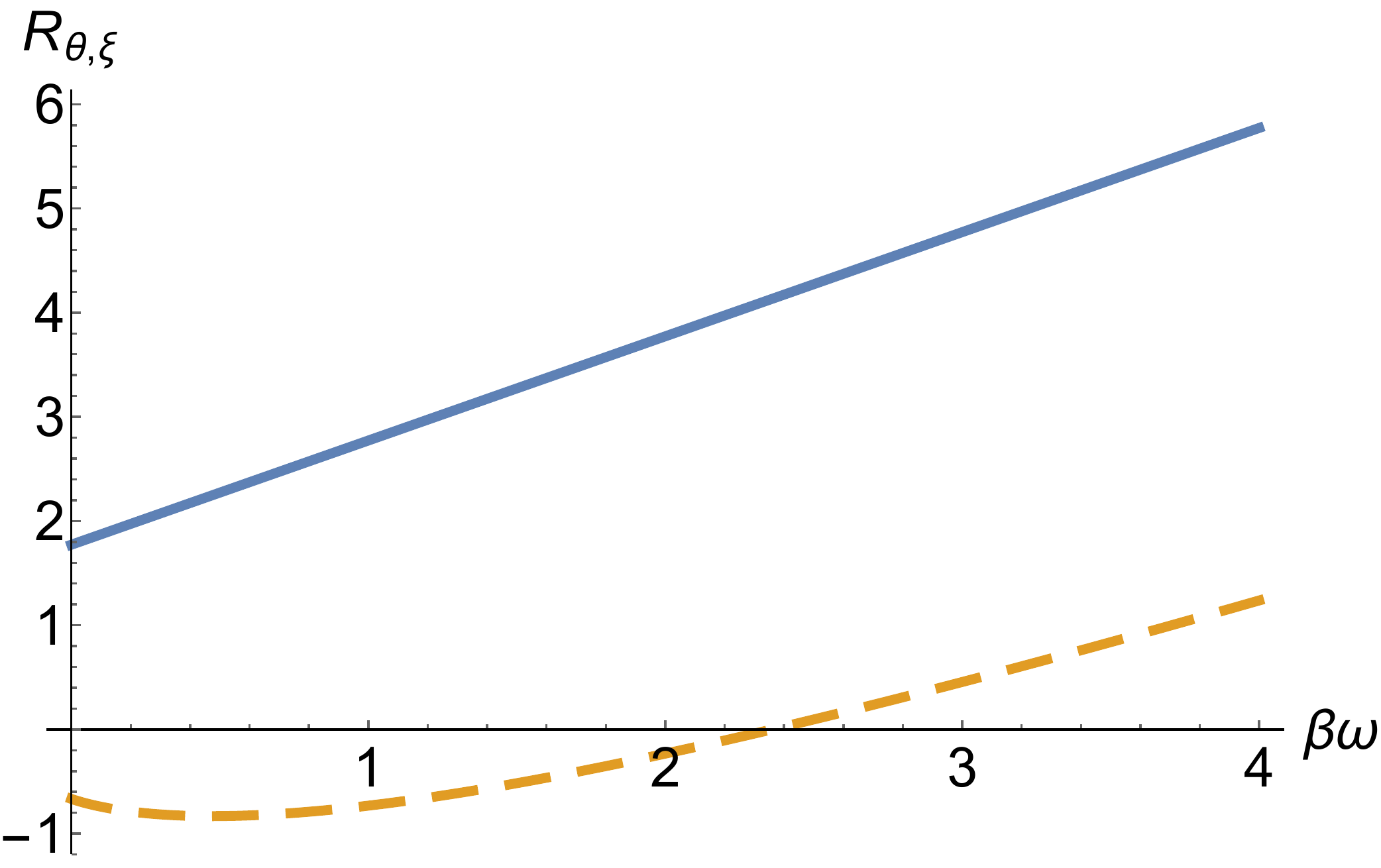}
  \caption{Ratios  (\ref{PHOTO}) for the photon thermal bulk contribution $R_\zeta$ blue-solid curve vs $\beta\omega$, 
  and for the photon thermal shear contribution $R_\eta$ orange-dashed curve vs $\beta\omega$, and 
  equal relaxation times $t_\zeta=t_\eta=\beta$ for fixed $\theta=\sigma=1$.}
    \label{fig_rate_photon}
  \end{center}
\end{figure}

As we noted in the dilepron rates above, the viscous corrections in leading order correspond to the
partinent insertions of $k_ik_j$ on the partonic lines as given in (\ref{001}-\ref{002}). These gradient insertions
break the isotropic character of the integrations with the typical integral structures

\be
&&q_iq_j\int {\rm cos}^2\theta +\frac{(\delta_{ij}-q_iq_j)}{2}\int (1-{\rm cos}^2\theta)\nonumber \\ 
&&=q_iq_j\int\frac{3\,{\rm cos}^2\theta-1}{2} +\delta_{ij}\int \frac{1-{\rm cos}^2\theta}{2}
\ee
where the scattering angles are defined as 

\be
&&{\rm cos}\,\theta_{1}=\frac{t+2EE_1}{2EE_1}\nonumber\\
&&{\rm cos}\,\theta_2=\frac{u+2EE_2}{2EE_2}\nonumber\\
&&{\rm cos}\,\theta_3=\frac{2(E_1+E_2-E)E-s}{2E(E_1+E_2-E)}
\ee
The angles $\theta_{1,2,3}$ refer to  the angles between particle $1,2,3$ and $\gamma$ flying along the z-direction,
in the process labeled as $1+2\rightarrow 3+\gamma$.
To proceed, we now make two kinematical approximations to simplify the integration analyses to follow.
The first is to follow (\ref{APPX}) and approximate the population factors by

\be
f_{ \pm \mu}(E_1)(1-f_{\pm \mu}(E_1))f_g(E_2)\approx&& e^{-\beta(E_1+E_2\mp \mu)}\nonumber\\
f_{\pm \mu}(E_1)(f_g)(1+f_g) \approx &&e^{-\beta(E_1+E_2\mp \mu)}\nonumber\\
\ee
and the second is to replace $E_{1,2}$ in ${\rm cos}(\theta_{1,2})$ by 
$E_{1,2}\approx (E_1+E_2)/2$. These approximations will allow us to extract explicit estimates 
for the viscous perturbative and non-perturbative effects. They will be tested against realistic 
hydrodynamical evolution of the rates in the future. 
With this in mind the viscous corrections to  the Compton (\ref{VCV}) 
and pair (\ref{VPV}) are respectively given by

\be
\label{FCP}
&&-G^{<}_{c}= t_H \partial_i \beta_j\left(\frac{ \delta_{ij}-\hat q_i\hat q_j}{2}G_c^{P}+ \frac{3\hat q_i\hat q_j-\delta_{ij}}{2}G_c^{T}\right)
\nonumber\\
&&-G^{<}_{p}= t_H \partial_i \beta_j\left(\frac{ \delta_{ij}-\hat q_i\hat q_j}{2}G_p^{P}+ \frac{3\hat q_i\hat q_j-\delta_{ij}}{2}G_p^{T}\right)
\nonumber\\
\ee
with the Compton kernels 

\be
\label{GC}
G_c^{P} \approx &&\frac{{\bf N}}{8(2\pi)^3E }\frac 1{e^{\beta(E-\mu)}-1}\int \frac {ds dt}s\, {|M_c(s,t)|^2}\nonumber \\ 
&&\times  \int_{\frac{s}{8E}}^{\infty} dx \left(\frac{2x+E}{e^{\beta(2x-\mu)}+1}-\frac{2xe^{\beta(2x-\mu)}}{(e^{\beta(2x-\mu)}+1)^2}\right)
\nonumber\\
&&+\mu \rightarrow -\mu\nonumber\\ 
G_c^{T} \approx && \frac{{\bf N} }{8(2\pi)^3E }\frac 1{e^{\beta(E-\mu)}-1}
\int  \frac{ds dt}s\, {|M_c(s,t)|^2}\nonumber \\ 
&&\times  \int_{\frac{s}{8E}}^{\infty} dx \biggl(\frac{2x+E-\frac{s}{E}+\frac{t^2+(s+t)^2}{8xE^2}}{e^{\frac{2x-\mu}{T}}+1} \nonumber \\
&&\qquad \qquad\,\,\,- \frac{2xe^{\beta(2x-\mu)}(1-\frac{s}{4Ex})^2}{(e^{\beta(2x-\mu)}+1)^2}\biggr)\nonumber\\
&&+\mu \rightarrow -\mu
\ee
and the pair production kernels

 \be
 \label{GP}
G_p^{P} \approx &&\frac{{\bf N} }{8(2\pi)^3E }\frac 1{e^{\beta E}-1}\int  \frac{ds dt}s \, {|M_p(s,t)|^2} \nonumber \\ 
&&\times  \int_{\frac{s}{8E}}^{\infty} dx\left( \frac{2x+E}{e^{\beta 2x}-1}+\frac{2xe^{\beta 2x}}{(e^{\beta 2x}-1)^2}\right)\nonumber\\ 
G_p^{T} \approx &&\frac{{\bf N} }{8(2\pi)^3E }\frac 1{e^{\beta E}-1}\int  \frac{ds dt}s\, {|M_p(s,t)|^2}\nonumber \\ 
&&\times  \int_{\frac{s}{8E}}^{\infty} dx \biggl(\frac{2x+E-\frac{s}{E}+\frac{t^2+(s+t)^2}{8xE^2}}{e^{\beta 2x}-1} \nonumber \\
&&\qquad \qquad\,\,\,+ \frac{2xe^{\beta 2x }(1-\frac{s}{4Ex})^2}{(e^{\beta 2x }-1)^2}\biggr)
\ee 
which are independent of the chemical potential $\mu$.  In both kernels in (\ref{GC}-\ref{GP}) the substitution 
(\ref{KMS1}) was performed  to recover the causal pre-factor required by the KMS condition. 
In terms of the hydrodynamical times, (\ref{FCP}) reads

\be
\label{FCPX}
&&-G^{<}_{c}= \left(\frac{ 2t_\zeta\theta-t_\eta\sigma}{2}G_c^{P}+ \frac{3t_\eta\sigma}{2}G_c^{T}\right)
\nonumber\\
&&-G^{<}_{p}=\left(\frac{ 2t_\zeta\theta-t_\eta\sigma}{2}G_p^{P}+ \frac{3t_\eta\sigma}{2}G_p^{T}\right) \nonumber\\
\ee
and the perturbative viscous correction to the photon rate takes the final form

\be
\label{FCPXX}
\frac{d^3\Gamma^{\rm Vp}_1}{d^3k}=\frac{-1}{(2\pi)^32E}(G^{<}_{c}+G^{<}_{p})_{\mu=0}
\ee
after setting $\mu=0$ in (\ref{GC}). In Appendix B, we explicit the contributions in (\ref{FCPXX}). The ratio of the 
leading shear and bulk thermal contributions in (\ref{FCPXX}) to the leading thermal perturbative photon contribution 
(\ref{P1}) are found to be

\be
\label{PHOTO}
R_\eta=\frac{d^3\Gamma^{Vp}_{\rm shear}}{d^3\Gamma^T_0}\approx &&
\left(E+\frac 1\beta\left(A_1-2A_2-\frac{1}{3}-\frac{7\zeta_3}{\pi^2}\right)\right)\,t_{\eta}\sigma \nonumber\\
R_\zeta=\frac{d^3\Gamma^{Vp}_{\rm bulk}}{d^3\Gamma^T_0}\approx &&\left(E+\frac 1\beta\left(\frac{2}{3}+\frac{14\zeta_3}{\pi^2}\right)\right)\, t_{\zeta}\theta
\ee
with $E=\omega=k_0$. 
Here $A_{1,2}$ are functions of $E$ defined  in (\ref{A1A2}),
that asymptote zero at large $E$ exponentially, and $\zeta_3$ referes to Riemann zeta function.  In Fig.~\ref{fig_rate_photon}
we show the ratios (\ref{PHOTO}) vs $\beta\omega$ with $\omega=E$, for $t_\zeta=t_\eta=\beta$ and fixed $\theta=\sigma=1$.
The thermal viscous corrections to the photon emissivities become linearly large at large $E=\omega$. Since (\ref{PHOTO}) 
were derived for large $E=\omega$, the low $E=\omega$ part of the curve receives additional corrections. Note that the
constants contributions were dropped from both the perturbative and viscous rates at large $E=\omega$.

\subsubsection{Thermal non-perturbative contributions}

To obtain the thermal non-perturbative corrections to the photon rates we proceed as in the case of 
the dilepton rates above, by expanding  (\ref{GC}) to order $\mu^2$ and then trading $\mu^2$  as in
(\ref{S1}). With this in mind, we have

\be
-G_c|_{\mu^2}=(\beta\mu)^2(G_1+G_2)
\ee
with

\be
G_1=&&\frac{{\bf N }}{16(2\pi)^3\beta E}n_B(1+n_B)(1+2n_B)\nonumber \\
&&\times \int \frac{dsdt}{s}|M_c(s,t)|^2\rm ln(1+e^{-\frac{s}{4E}})\nonumber\\
G_2=&&\frac{{\bf N}}{16(2\pi)^3\beta E}n_B\int \frac{dsdt}{s}|M_c(s,t)|^2\frac{e^{\frac{\beta s}{4}}}{(1+e^{\frac{\beta s}{4}})^2}\nonumber\\
\ee
The associated non-perturbative photon rate is

\be
\label{P4X}
-\frac {\beta^2}{(2\pi)^32E}{\left<(gA_4)^2\right>}(G_1+G_2)_{\mu=0}
\ee
which corresponds to the leading soft gluon insertion to the Compton process. We now note that eventough 
the perturbative annihilation process $q\bar q\rightarrow \gamma$ is kinematically forbidden, its counterpart in 
the presence of external fields $(gE), (gB)$ is not. This contribution can be obtained from 
the $\gamma^*\rightarrow q\bar q$ process in (\ref{P4}) after suitably multiplying by $q^2$ and then taking
$q^2\rightarrow 0$ to recover the photon point through the dilepton-photon identity

\be
\label{EG}
q_0\frac{d^3\Gamma}{d^3q}=\frac{3\pi}{2\alpha \mathbb B}\lim_{q^2\to 0}\left(q^2\frac{d^4{\mathbb R}}{d^4q}\right)
\ee
The outcome combined with (\ref{P4X}) gives

\be
\label{P4XY}
\frac{d^3\Gamma^{Tn}_0}{d^3k}=&&\frac{\alpha}{2\pi^2 E}\frac 1{e^{\beta E}-1}
\left(\frac 1{4\pi}\,d_F\sum_f \hat e_f^2\right)\nonumber\\
&&\times\left( \frac 1{6}\left<(gE)^2\right>
- \frac 1{3}\left<(gB)^2\right>\right)\nonumber\\
&&\times \left(\frac{\beta}{|q|}\right)\left(n_+(1-n_+)-n_-(1-n_-)\right)\biggr)\nonumber\\
&&-\frac {\beta^2}{(2\pi)^32E}{\left<(gA_4)^2\right>}(G_1+G_2)_{\mu=0}
\ee
The explicit forms of $G_{1,2}$ follow the same analysis detailed in Appendix B. 
This is our final result for the thermal and non-perturbative contributions for the photon emissivity, in leading mass dimensions.

\subsubsection{Viscous non-perturbative contributions}

The viscous non-perturbative contributions of the type $\left<(gA_4)^2\right>$ are readily obtained by expanding (\ref{GC})
in powers of $\mu^2$ and using the identification $\mu^2\rightarrow -\left<(gA_4)^2\right>$ as we discussed above. 
We note that the $\mu$ dependence drops out of the pair production rate (\ref{GP}). More specifically, we have for the
Compton contribution to second order in $\mu^2$

\be
\label{X0X0X0}
G_c^{P}|\mu^2=(\beta\mu)^2G_3\qquad\qquad G_c^{T}|_{\mu^2}=(\beta\mu)^2G_4
\ee
with 

\be
&&G_3=(1+n_B(E))(1+2n_B(E))\,G_c^{P}\nonumber \\ &&+\frac{{\bf N}}{8(2\pi)^3E }\,n_B(E)\,
\int  \frac{ds dt}s |M_c(s,t)|^2\nonumber \\  
&&\times \int_{\frac{s}{8E}}^{\infty}\biggl( (2x+E)f(1-f)\nonumber\\
&&-2xf(1-f)(6f^2-6f+1)\biggr)\nonumber\\\nonumber\\
&&G_4=(1+n_B(E))(1+2n_B(E))\,G_c^{T}\nonumber \\ &&+\frac{{\bf N} }{8(2\pi)^3E }\,n_B(E)\,
\int  \frac{ds dt}s |M_c(s,t)|^2\nonumber \\ 
&&\times\int_{\frac{s}{8E}}^{\infty}\biggl(\left(2x+E-\frac{s}{E}+\frac{t^2+(s+t)^2}{8xE^2}\right)f(1-f)\nonumber \\ 
&&-2x\left(1-\frac{s}{4Ex}\right)^2f(1-f)(6f^2-6f+1)\biggr)
\ee
Here we have set $f=f(2x)=1/(e^{\beta 2x}+1)$. 
The non-perturbative viscous corrections to the photon rates are

\be
\label{VNX}
&&\frac{d^3\Gamma^{Vn}_1}{d^3k}=\frac{-1}{(2\pi)^32E}G^{<}(E)=\nonumber \\ &&+\frac{\alpha}{2\pi^2E}\frac{1}{e^{\beta E}-1}\left(\frac 1{4\pi}\,d_F\sum_f \hat e_f^2\right)\, \frac {\beta^2}{\pi E}\nonumber\\
&&\times \left(  \frac 1{6}\left<(gE)^2\right>
-\frac 1{3}\left<(gB)^2\right>\right)\nonumber\\
&&\biggl( \frac{2 t_\zeta\theta-t_\eta\sigma}{2}(n_B F_1+n_B(1+n_B)F_2+n_B^2(1+n_B)F_3)\biggr.\nonumber \\
&&\biggl.+\frac{3t_\eta\sigma}{2}(n_B \tilde F_1+n_B(1+n_B)\tilde F_2+n_B^2(1+n_B)\tilde F_3)\biggr) \nonumber\\
&&-\frac {\beta^2}{(2\pi)^32E}{\left<(gA_4)^2\right>}\left(\frac{ 2t_\zeta\theta-t_\eta\sigma}{2}G_3+ \frac{3t_\eta\sigma}{2}G_4\right)\nonumber\\
\ee
The first contribution stems again from the $q^2\rightarrow 0$ of the non-perturbative viscous contribution for the
$\gamma^*\rightarrow q\bar q$ rate in (\ref{P2X2}) using (\ref{EG}), and the last  contribution follows from (\ref{X0X0X0}) after 
inserting it in (\ref{FCPX}),  and combining it  with (\ref{FCPXX}) following the
substitution $\mu^2\rightarrow -\left<(gA_4)^2\right>$. 
Again, the explicit forms of $G_{3,4}$ follow the same analysis detailed in Appendix B. 
(\ref{VNX}) is our final result for the viscous and non-perturbative corrections
to the photon rates in leading mass dimensions.

\section{conclusion}

We have provided a general framework for analyzing near-equilibrium hydrodynamical corrections to the
photon and dilepton emissivities in QCD. Assuming that the emission times are short in comparison to the 
hydrodynamical evolution times, we have developed the rates by expanding the evolving fluid density matrix in 
derivatives of the fluid gradients. In leading order, the electromagnetic rates get corrected by bulk and shear viscous 
contributions in the form of   Kubo-like response functions involving the energy-momentum tensor.

We have analyzed the viscous corrections in a hadronic fluid below the QCD transition temperature for both
the photon and dilepton emissivities. A simple estimate of the photon rate using vector dominance in the chiral limit
shows that the bulk viscosity corrections are much larger than the shear viscosity corrections for about all 
frequencies. The former are still sizable in the late stages of the hadronic evolution.
These observations are interesting to check in a full hydrodynamical
analysis of the photon emissivities at present colliders. Similar corrections were also shown to occur in the dilepton
emission rates. 

We have also analyzed the viscous corrections in a strongly coupled quark gluon plasma (sQGP)  for temperatures
higher but close to the  transition temperature, as  probed by current colliders. The non-perturbative character of the
sQGP is developed by correcting the thermal perturbative rates with soft gluonic insertions in the form of gluonic 
operators of increasing mass dimensions, in the spirit of the OPE expansion for the QCD vacuum correlation functions.
The partonic thermal bulk viscous corrections to the dilepton and photon rates are observed to be more sizable than their shear
counterparts with increasing dilepton and photon energies. 
As our calculations were carried at finite chemical potential $\mu$ which was traded by an expansion with $igA_4$,
they also provide for the viscous corrected electromagnetic rates at finite chemical potential as well.

The shortcomings of our analysis stem from our decoupling approximation that the hydrodynamical gradients 
decorrelate on time scales that are larger than the electromagnetic emission times, and also our also assessment of
only the leading gradient corrections. To improve on this, looks at this stage formidable. This notwithstanding, 
the present viscous corrections to the hadronic and partonic emissivities can and should be assessed in current
analyses with hydrodynamical base evolution. In particular their effects on the currently reported photon flow
\cite{VISCO,RALF,LEEX}.

\section{Acknowledgements}

We thank Jean-Francois Paquet for a discussion.
This work was supported by the U.S. Department of Energy under Contract No.
DE-FG-88ER40388.

\section{Appendix A: Background field analysis}

\begin{figure}[b]
  \begin{center}
  \includegraphics[width=8cm]{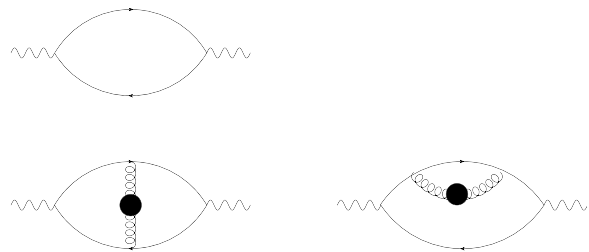}
  \caption{The thermal $JJ$ correlation function in Euclidean space. The soft gluon insertions 
   of the type $(gA_4)^2, (gE)^2, (gB)^2$ are indicated by a blob.}
  \label{fig_loop}
  \end{center}
\end{figure}

In this Appendix we briefly outline how to correct the Wightman function for the $JJ$ correlator
using the background field method as initially discussed in~\cite{OPE} and illustraded in
Fig.~\ref{fig_loop}. The soft gluon corrections are indicated by a blob. For the case of the  leading
$(gA_4)^2$ insertion discussed above, this construction relies directly on Feynman diagrams
in the background field method, rather than the observation that $A_4$ plays the role of a colored 
chemical potential, and therefore can be traded by a real chemical potential as we discussed above.
It also shows how the $(gE)^2$ and $(gB)^2$ corrections are obtained.  Throughout this appendix 
the analysis is in Euclidean space and we will set $A_4=A_0$.

Using the Fock-Schwinger gauge for the background fields, we can explicitly re-write
the gauge fields as an expansion in increasing covariant derivatives of the field strengths,

\be
\label{APP0}
&&A_0=A_0+x^{i}D_{i}A_0+...\nonumber\\
&&A_i=x^{j}F_{ji}+\sum_{k=1}\frac{k+2}{k!}x^{j}x^{j_1}..x^{j_{k}}D_{j_1}..D_{j_k}F_{ji}
\ee
The fermion propagator in these external fields takes the form

\be
\label{APPX}
S(q)=\int d^4x \,e^{-iq\cdot x}\left<x\biggl|\frac{1}{-i\gamma \cdot D}\biggr|0\right>
\ee

For simplicity, consider first the presence of a magnetic field $B$ by limiting the external
gauge field in $D$ to $A_i$. Expanding (\ref{APPX}) to first order in $B$, we have

\be
&&S_1(q)=\nonumber \\ 
&&-g\int \frac{d^4q_1d^4q_2}{(2\pi)^4(2\pi)^4}\frac{q_1\cdot \gamma}{q_1^2}F_{ij}\gamma^{j}\frac{q_2\cdot \gamma}{q_2^2}i\frac{\partial\delta^4(q_1-q_2)}{\partial (q_1-q_2)^i}e^{iq_1\cdot x}\nonumber\\
\ee
using $\frac{\partial\delta^4(q_1-q_2)}{\partial (q_1-q_2)^i}=-\frac{\partial\delta^4(q_1-q_2)}{\partial q_2^i}$
we obtain

\be
&&S_1(q)=-igS_0(q)F_{ij}\gamma^j\partial_{i}S_0(q)\nonumber\\
&&S_0(q)=\frac{\slashed{q}}{q^2}
\ee
After a simple reduction we have to first order in $B$

\be
S_1[B]=-i\frac{gF_{ij}}{q^4}\gamma^{i}\slashed{q}\gamma^{j}
\ee
The second order correction in $B$ follows by expanding further in (\ref{APPX}), giving the following contribution

\be
&&-g^2S_0(q_1)F_{ij}\gamma^jS_0(q_2)F_{kl}\gamma^{l}S_0(q_3)\nonumber \\
&& \times \partial_{q_2-q_3}\delta(q_2-q_3)\partial_{q_1-q_2}\delta(q_1-q_2)
\ee
which can be reduced by  first partially integrating with respect to $q_3$, and then  partially integrating with respect to $q_2$,  
to obtain

\be
\label{APP8}
S_2[B]=-\frac{g^2F_{ij}F_{kl}}{q^6}\slashed{q}\left(\gamma^{j}\gamma^{k}\gamma^{i}\gamma^{l}-\frac{4q_i\gamma_j}{q^2}\gamma^{k}\slashed{q}\gamma^{l}\right)
\ee
Color-spin averaging (\ref{APP8}) using 

\be
\left<F_{kl}F_{mn}\right>=\frac 18\frac 13 {\left<B^2\right>}(\delta_{km}\delta_{ln}-\delta_{kn}\delta_{lm})\nonumber
\ee
leads to the $B^2$ correction to the fermionic propagator

\be
\left<S_2[B]\right>=\frac{2\left<(gB)^2\right>}{3q^6}\left(q_0\gamma_0-\frac{q_0^2}{q^2}\slashed{q}\right)
\ee

Using a similar reasoning as for the magnetic field $B$, we can seek the corrections in $A_0$ and $E$ 
to second order. The results are for the electric field

\be
&&S_1[E]=-i\frac{gD_{i}A_0}{q^4}\left(\gamma^{i}\slashed{q}\gamma^{0}-\frac{4q_0 q_i}{q^2}\slashed{q}\right)\nonumber\\
&&\left<S_2[E]\right>=\frac{\left<(gE)^2\right>}{8q^6}(\slashed{q}f_1+q_0\gamma_0f_2)\nonumber\\
&&f_1=\frac{1}{3}+\frac{28q_0^2}{3q^2}-\frac{16q_0^2|\vec{q}|^2}{q^4}\nonumber\\
&&f_2=-\frac{20}{3}+ \frac{8|\vec{q}|^2}{3q^2}
\ee
and for $A_0$

\be
&&S_1[A_0]=-\frac{gA_0}{q^4}\slashed{q}\gamma_0\slashed{q}\nonumber\\
&&S_2[A_0]=\frac{g^2A_0^2} {q^6}\slashed{q}\gamma_0\slashed{q}\gamma_0\slashed{q}
\ee

The leading magnetic correction to the Wightman function for the $JJ$ correlator in Euclidean space  is

\be
\label{EUCLIDB}
&&\left<\Gamma_{\mu \mu}^E(q;[B])\right>=e^2\left<(gB)^2\right>(I_1+I_2)\nonumber\\
&&I_1=-\frac{16}{3}\sum_{q_1+q_2=q}\frac{|\vec{q_1}|^2\omega_1\omega_2-\omega_1^2 \vec{q_1}\cdot \vec{q_2}}{q_1^8q_2^2}\nonumber\\
&&I_2=-2\sum_{q_1+q_2=q} \frac{\omega_1\omega_2+\frac{\vec{q_1}\cdot \vec{q_2}}{3}}{q_1^4q_2^4}
\ee
with $q=(\omega, \vec q)$ and the short hand notation 

\be
\sum_q\equiv \sum_{\omega_n=\pi T(2n+1)}\int \frac{d^3\vec{q}}{(2\pi)^3}\nonumber
\ee

The leading $A_0$ correction is

\be
\label{EUCLIDA}
&&\left<\Gamma_{\mu \mu}^E(q;[A])\right>=e^2\left<(gA_0)^2\right>(I_3+I_4)\nonumber\\
&&I_3= -8\nonumber \\
&&\times \sum_{q_1+q_2=q}\frac{\omega_1\omega_2(\omega_1^2-3|\vec{q_1}|^2)+(3\omega_1^2-|\vec{q_1}|^2)\vec{q_1}\cdot\vec{q_2}}{q_1^6q_2^2}\nonumber\\
&&I_4=-4\nonumber \\ 
&&\times\sum_{q_1+q_2=q} \frac{(\omega_1^2-|\vec{q_1}|^2)(\omega_2^2-|\vec{q_2}|^2)+4\omega_1\omega_2\vec{q_1}\cdot{\vec{q_2}}}{q_1^4q_2^4}\nonumber\\
\ee
Here $I_4$ follows from the soft insertion in the second contribution (bottom) of Fig.~\ref{fig_loop},
and $I_3$ follows from the soft insertion in the first contribution (bottom) of Fig.~\ref{fig_loop}.

If we set  the external momentum $q=(q_0,\vec 0)$, then (\ref{EUCLIDB}-\ref{EUCLIDA}) can be reduced to

\be
\label{EUCLIDALL}
&&I_1=\frac{32}3\left(2q_0I_{31}^0+I_{31}^{00}+q_0^2I_{41}^{00}-I_{40}^{00}\right)\nonumber\\
&&I_2=\frac{-1}{3}\left(2I_{12}-q_0^2I_{22}+4q_0I_{22}^0+4I_{22}^{00}\right)\nonumber\\
&&I_3=4\left(-3q_0I_{21}^0+4q_0I_{31}^{000}+2I_{21}^{00}-I_{11}\right)\nonumber\\
&&I_4=2\left(+2q_0^2I_{22}^{00}-q_0^2I_{12}+2q_0I_{12}^0+I_{02}\right)
\ee
with the following notation  

\be
I_{mn}^{\mu_1...\mu_j}\equiv \sum_k\frac{k^{\mu_1}...k^{\mu_j}}{k^{2m}(k+q)^{2n}}
\ee
some useful properties of these integrals can be found in~\cite{Hansson:1987un}.  
In particular, we have the identity

\be
&&2I_{12}-q_0^2I_{22}=4q_0I_{22}^0+4I_{22}^{00}
\ee
The magnetic contribution $I_2$
in (\ref{EUCLIDALL}) diverges in the infrared at zero temperature. This contribution can
be  reabsorbed in the definition of $m\bar q q$ at zero temperature. This will be assumed at finite
temperature as well. Since the chiral condensate vanishes in the partonic phase, we will set this self-energy 
type contribution to zero. The same will be assumed for the analogue electric contribution. With this in mind, 
the electric and magnetic contributions following from the soft gluon insertions contribute to the
$JJ$ correlator as

\be
\left<\Gamma_{\mu\mu}^E(q;[E,B])\right>=&&-4e^2\left<(gE)^2\right>\left(2I_{12}-q_0^2I_{22}\right)\nonumber\\
&&+\frac {4e^2}3 \left(\left<(gE)^2\right>+\left<(gB)^2\right>\right)\nonumber\\
&&\times \left(2I_{12}-q_0^2I_{22}+4q_0I_{22}^0+4I_{22}^{00}\right)\nonumber\\
\ee
The asymmetry between the electric and magnetic field is due to the breaking of Lorentz invariance introduced by the
heat bath. The Matsubara summation in the I-integrals reduce to a zero temperature plus a finite temperature part
through the use of

\be
&&\sum_{\omega_n=\pi T(2n+1)}\beta F(\omega_n)=\nonumber\\
&&\int_{-i\infty+i0}^{+i\infty+i0}
\frac{dz}{2i\pi}\left(\frac 12-f(z)\right)\left(F(z)+F(-z)\right)
\ee
with  $f(z)$ a thermal Fermi distribution. The analytical continuation of the Euclidean correlator to its Minkowski counterpart  follows through the discontinuity

\be
\Gamma^{\mu<}_{\mu}(\omega)=\frac{1}{i(e^{\beta \omega}-1)}\,\rm {Disc}\,\Gamma_{\mu \mu}^E(q_0\rightarrow -i(\omega\pm i\epsilon))
\ee
The results for the $(gA_0)^2$ insertion
are in complete agreement with those obtained using a chemical potential and then 
the substitution~(\ref{S1}). The results for the $(gE)^2, (gB)^2$ insertions correspond to the substitution (\ref{S2}).

\section{Appendix B: Leading thermal viscous photon correction}

In this Appendix we explicit the calculations leading to (\ref{PHOTO}). We start by performing the 
integrals

\be
&&\int \frac{dt}{s}(t^2+(s+t)^2)|M_c(s,t)|^2=128\pi^2\alpha \alpha_s\nonumber \\ 
&&\times \left(-\frac{2 m_T^6}{3s}+m_T^4 +m_T^2 s-\frac{2 s^2}{3}+s^2{\rm ln} \left(\frac{s-m_T^2}{m_T^2}\right)\right)\nonumber\\
&&\int \frac{dt}{s}(t^2+(s+t)^2)|M_p(s,t)|^2=128\pi^2\alpha \alpha_s\nonumber \\ 
&&\times\left(\frac{8 m_T^6}{3s}-4m_T^4 +8m_T^2 s-\frac{10 s^2}{3}+2s^2{\rm ln} \left(\frac{s-m_T^2}{m_T^2}\right)\right)\nonumber\\
\ee
Much like the leading perturbative thermal contribution (\ref{P1}), the viscous thermal contributions
are also infrared sensitive. The leading singularities are  logarithmic. For the Compton $G_c$ 
and the pair $G_p$ amplitudes they are

\be
G_c^P\approx && \frac{{\bf N} }{8(2\pi)^3 \beta E}\frac {128\pi^2\alpha \alpha_s}{e^{\beta E}-1} \nonumber\\
&&\times \int_{2m_T^2}^{\infty} ds 
{\rm ln}\left(\frac{s-m_T^2}{m_T^2}\right)
\sum_{n=1}^{\infty}\frac{(-1)^{n+1}e^{-\frac{ns\beta}{4E}}}{n}\nonumber\\
&&\times\left(E-\frac 1\beta\left(1-\frac{1}{n}\right)\left(1+\frac{ns\beta}{4E}\right)\right)\nonumber\\
G_c^T \approx && \frac{{\bf N}  }{4(2\pi)^3 \beta E}\frac {128\pi^2\alpha \alpha_s}{e^{\beta E}-1}\nonumber\\
&&\times \int_{2m_T^2}^{\infty} ds
{ \rm ln}\left(\frac{s-m_T^2}{m_T^2}\right)
\sum_{n=1}^{\infty}\frac{(-1)^{n+1}e^{-\frac{ns}{4ET}}}{2n}\nonumber\\
&&\times\left(E-\frac 1\beta\left(1-\frac{1}{n}\right)\left(1+\frac{ns\beta}{4E}\right)+\frac{s}{E}(n-1)\right)\nonumber \\ 
&&+\int_{\frac{s}{8E}}^{\infty} dx(-1)^{n+1}e^{-2n\beta x}
\left(\frac{s^2}{4E^2(x+\frac{E}{2})}-n\frac{s^2}{8E^2x}\right)\nonumber\\
\ee
and

\be
G_p^P\approx && \frac{{\bf N} }{8(2\pi)^3\beta E}\frac {128\pi^2\alpha \alpha_s}{e^{\beta E}-1}
\nonumber\\&&\times \int_{2m_T^2}^{\infty} ds {\rm ln}\left(\frac{s-m_T^2}{m_T^2}\right)\nonumber \\
&&\times \sum_{n=1}^{\infty}\frac{e^{-\frac{ns\beta}{4E}}}{n}\left(E+\frac 1\beta\left(1+\frac{1}{n}\right)\left(1+\frac{ns\beta}{4E}\right)\right)\nonumber\\
G_p^T \approx &&\frac{{\bf N}  }{4(2\pi)^3\beta E}\frac {128\pi^2\alpha \alpha_s}{e^{\beta E}-1}\nonumber\\
&&\times \int_{2m_T^2}^{\infty} ds {\rm ln}\left(\frac{s-m_T^2}{m_T^2}\right)\nonumber \\ 
&&\times  \sum_{n=1}^{\infty}\frac{e^{-\frac{ns\beta}{4E}}}{2n}\nonumber\\
&&\times \left(E+\frac 1\beta\left(1+\frac{1}{n}\right)\left(1+\frac{ns\beta}{4E}\right) -\frac{s}{E}(n+1)\right)\nonumber \\ 
&&+\int_{\frac{s}{8E}}^{\infty} dx e^{-2n\beta x}\left(\frac{s^2}{4E^2(x+\frac{E}{2})}+n\frac{s^2}{8E^2x}\right)\nonumber\\
\ee
For a small infrared cutoff $m_T$, we now define the useful integrals

\be
&&\int_{2m_T^2} ds s^k {\rm ln}\left(\frac{s-m_T^2}{m_T^2}\right)\,e^{-\frac{ns\beta}{4E}}\nonumber \\ 
&&=k!\,\left(\frac{4E}{n\beta}\right)^{k+1}{\rm ln}\left(\frac{4E}{\beta m_T^2}\right)+{\mathbb C}
\ee
and

\be
&&\int_{2m_T^2} ds s^k{\rm ln}\left(\frac{s-m_T^2}{m_T^2}\right)\int_{\frac{s}{8E}}dx \frac{e^{-2n\beta x}}{x+a}\nonumber \\ 
&&=k!\,{\rm ln}\left(\frac{4E}{\beta m_T^2}\right)\int_{1}^{\infty}\left(\frac{4E}{un\beta}\right)^{k+1}\frac{e^{-2n\beta a(u-1)}}{u}du +{\mathbb C}\nonumber\\
\ee
Collecting the above results yield for the Compton and pair amplitudes

\be
G_c^{P}=&&\frac{{\bf N} }{2(2\pi)^3\beta^2 }\frac {128\pi^2\alpha \alpha_s}{e^{\beta E}-1}
\biggl({\rm ln}\left(\frac{4E}{\beta m_T^2}\right)\nonumber \\
 &&\times\sum_{n=1}\frac{(-1)^{n-1}}{n^2}\left(E-\frac 2\beta\left(1-\frac{1}{n}\right)\right)+{\bf C_1}\biggr)\nonumber\\ 
G_c^{T}=&&\frac{{\bf N} }{2(2\pi)^3 \beta^2}\frac {128\pi^2\alpha \alpha_s}{e^{\beta E}-1}
\biggl({\rm ln}\left(\frac{4E}{\beta m_T^2}\right)\nonumber \\  
&&\times\biggl(\sum_{n=1}\frac{(-1)^{n-1}}{n^2}
\biggl(E+\frac 2\beta\left(1-\frac{1}{n}\right)\nonumber \\ 
&&+\frac{8}{n\beta}\left(I_4\left(n,\frac{E}{2}\right)-\frac{n}{2}I_4(n,0)\right)\biggr)\biggr)+{\bf C_2}\biggr)\nonumber\\
\ee
and 

\be
G_p^{P}=&&\frac{{\bf N}}{2(2\pi)^3 \beta^2}\frac {128\pi^2\alpha \alpha_s}{e^{\beta E}-1}
\biggl({\rm ln}\left(\frac{4E}{\beta m_T^2}\right)\nonumber \\ 
&&\times\sum_{n=1}\frac{1}{n^2}\left(E+\frac 2\beta \left(1+\frac{1}{n}\right)\right)+{\bf C_3}\biggr)\nonumber\\
G_p^{T}=&&\frac{{\bf N} }{2(2\pi)^3\beta^2 }\frac {128\pi^2\alpha \alpha_s}{e^{\beta E}-1}
\biggl({\rm ln}\left(\frac{4E}{\beta m_T^2}\right)\nonumber \\  
&&\biggl(\times\sum_{n=1}\frac{1}{n^2}\biggl((E-\frac 2\beta\left(1+\frac{1}{n}\right)\nonumber \\ 
&&+\frac{8}{n\beta}\left(I_4\left(n,\frac{E}{2}\right)+\frac{n}{2}I_4(n,0)\right)\biggr)\biggr)+{\bf C_2}\biggr)\nonumber\\
\ee
where we have defined

\be
I_k(n,a)=&&\int_1^{\infty}\frac{e^{-2n\beta a(u-1)}}{u^k}du\nonumber\\
I_k(n,0)=&&\frac{1}{k+1}
\ee
Here ${\bf C}_{1,2,3,4}$ are constants. 
After further simplifications we finally obtain

\be
&&G_c^P=\frac{2\pi \alpha\alpha_s{\bf N}}{3\beta^2(e^{\beta E}-1)}
\left(E-\frac 2\beta\left(1-\frac{9\zeta_3}{\pi^2}\right)\right){\rm ln}\left(\frac{4E}{\beta m_T^2}\right)\nonumber\\
&&G_p^P=\frac{4\pi \alpha\alpha_s{\bf N}}{3\beta^2(e^{\beta E}-1)}
\left(E+\frac 2\beta\left(1+\frac{6\zeta_3}{\pi^2}\right)\right){\rm ln}\left(\frac{4E}{\beta m_T^2}\right)\nonumber\\
&&G_c^T=\frac{2\pi \alpha\alpha_s{\bf N}}{3\beta^2(e^{\beta E}-1)}
\left(E+\frac 2\beta A_1\right){\rm ln}\left(\frac{4E}{\beta m_T^2}\right)\nonumber\\
&&G_p^T=\frac{2\pi \alpha\alpha_s{\bf N}}{3\beta^2(e^{\beta E}-1)}
\left(E-\frac 2\beta A_2\right){\rm ln}\left(\frac{4E}{\beta m_T^2}\right)
\ee
with the E-dependent functions

\be
\label{A1A2}
&&A_1=1-\frac{9\zeta_3}{\pi^2}-\frac{48}{\pi^2}\sum_{n=1}\frac{(-1)^{n}}{n^3}\left(I_4\left(n,\frac{E}{2}\right)-\frac{n}{2}I_4(n,0)\right)
\nonumber\\
&&A_2=1+\frac{6\zeta_3}{\pi^2}-\frac{24}{\pi^2}\sum_{n=1}\frac{1}{n^3}\left(I_4\left(n,\frac{E}{2}\right)+\frac{n}{2}I_4(n,0)\right)\nonumber\\
\ee
Here $\zeta_3\equiv \zeta(3)\approx 1.202$ refers to Riemann zeta function.

 \vfil

\end{document}